\documentclass[preprint2]{aastex}
\usepackage{amsmath}
\usepackage{booktabs}
\usepackage{longtable}
\usepackage{xspace}
\usepackage[T1]{fontenc}
\usepackage{apjfonts}
\usepackage{natbib}
\usepackage{graphicx}

\newcommand{\ms}{\ensuremath{\rm m\,s^{-1}}}

\newcommand{\gcmc}{\ensuremath{\rm g\,cm^{-3}}}
\newcommand{\gcc}{\gcmc}

\newcommand{\vsini}{\ensuremath{v \sin{i}}}

\newcommand{\rsun}{\ensuremath{R_\sun}}
\newcommand{\msun}{\ensuremath{M_\sun}}

\newcommand{\rstar}{\ensuremath{R_\star}}
\newcommand{\mstar}{\ensuremath{M_\star}}

\newcommand{\rhostar}{\ensuremath{\rho_\star}}

\newcommand{\rpl}{\ensuremath{R_{\rm p}}}
\newcommand{\mpl}{\ensuremath{M_{\rm p}}}

\newcommand{\rhopl}{\ensuremath{\rho_{\rm p}}}

\newcommand{\rearth}{\ensuremath{R_\earth}}
\newcommand{\mearth}{\ensuremath{M_\earth}}
\newcommand{\fearth}{\ensuremath{F_\earth}}

\newcommand{\msini}{\ensuremath{\mpl\sin{i}}}

\newcommand{\rprs}{\ensuremath{\frac{\rpl}{\rstar}}}
\newcommand{\mpms}{\ensuremath{\frac{\mpl}{\mstar}}}

\newcommand{\me}{\ensuremath{9.1\pm1.0}}
\newcommand{\mb}{\ensuremath{358\pm12}}
\newcommand{\md}{\ensuremath{13.6\pm2.0}}
\newcommand{\mcsini}{\ensuremath{416\pm16}}
\newcommand{\pere}{\ensuremath{0.78961\pm0.00001}}
\newcommand{\perb}{\ensuremath{4.15912\pm0.00001}}
\newcommand{\perd}{\ensuremath{9.0304\pm0.0003}}
\newcommand{\perc}{\ensuremath{596\pm2}}
\newcommand{\tte}{\ensuremath{2146.7641\pm0.0007}}
\newcommand{\ttb}{\ensuremath{2149.9785\pm0.0001}}
\newcommand{\ttd}{\ensuremath{2155.308\pm0.001}}
\newcommand{\ttc}{\ensuremath{1162\pm5}}
\newcommand{\secwe}{\ensuremath{0.01\pm0.13}}
\newcommand{\seswe}{\ensuremath{0.07\pm0.13}}
\newcommand{\secwb}{\ensuremath{0.009\pm0.03}}
\newcommand{\seswb}{\ensuremath{0.04\pm0.04}}
\newcommand{\secwd}{\ensuremath{-0.01\pm0.06}}
\newcommand{\seswd}{\ensuremath{0.03\pm0.08}}
\newcommand{\secwc}{\ensuremath{-0.40\pm0.04}}
\newcommand{\seswc}{\ensuremath{-0.35\pm0.06}}
\newcommand{\stellarmass}{\ensuremath{1.00\pm0.05}}
\newcommand{\ee}{0.06}
\newcommand{\eb}{0.011}
\newcommand{\ed}{0.025}
\newcommand{\ec}{\ensuremath{0.28\pm0.02}}
\newcommand{\rhoe}{\ensuremath{9.22\pm1.06}}
\newcommand{\rhob}{\ensuremath{1.02\pm0.02}}
\newcommand{\rhod}{\ensuremath{1.63\pm0.23}}
\newcommand{\re}{\ensuremath{1.76\pm0.04}}
\newcommand{\rb}{\ensuremath{12.47\pm0.22}}
\newcommand{\rd}{\ensuremath{3.59\pm0.07}}

\makeatletter

\begin{document}

\pagenumbering{arabic}

\title{New Insights on Planet Formation in WASP-47 from a Simultaneous Analysis of Radial Velocities and Transit Timing Variations}
\author{
Lauren~M.~Weiss\altaffilmark{1,9},
Katherine~M.~Deck\altaffilmark{2},
Evan~Sinukoff\altaffilmark{3,10},
Erik~A.~Petigura\altaffilmark{2,11},
Eric~Agol\altaffilmark{4,5},
Eve~J.~Lee\altaffilmark{6},
Juliette~C.~Becker\altaffilmark{7},
Andrew~W.~Howard\altaffilmark{2},
Howard Isaacson\altaffilmark{6},
Ian J.~M.~Crossfield\altaffilmark{8},
Benjamin~J.~Fulton\altaffilmark{3},
Lea~Hirsch\altaffilmark{6}
Bj{\"o}rn Benneke\altaffilmark{2}}

\altaffiltext{1}{Institut de Recherche sur les Exoplan\`{e}tes, Universit\'{e} de Montr\'{e}al, Montr\'{e}al, QC, Canada}
\altaffiltext{2}{California Institute of Technology, Pasadena, CA, USA}
\altaffiltext{3}{Institute for Astronomy, University of Hawai`i at M\={a}noa, Honolulu, HI, USA} 
\altaffiltext{4}{Astronomy Department, University of Washington, Seattle, WA, USA}
\altaffiltext{5}{NASA Astrobiology Institute Virtual Planet Laboratory, Seattle, WA, USA} 
\altaffiltext{6}{Astronomy Department, University of California, Berkeley, CA, USA}
\altaffiltext{7}{Astronomy Department, University of Michigan, Ann Arbor, MI, USA}
\altaffiltext{8}{Astronomy Department, University of California, Santa Cruz, CA, USA}

\altaffiltext{9}{Trottier Fellow}
\altaffiltext{10}{NSERC Fellow}
\altaffiltext{11}{NASA Hubble Fellow}

\begin{abstract}
Measuring precise planet masses, densities, and orbital dynamics in individual planetary systems is an important pathway toward understanding planet formation.  The WASP-47 system has an unusual architecture that motivates a complex formation theory.  The system includes a hot Jupiter (``b'') neighbored by interior (``e'') and exterior (``d'') sub-Neptunes, and a long-period eccentric giant planet (``c'').  We simultaneously modeled transit times from the Kepler K2 Mission and 118 radial velocities to determine precise masses, densities, and Keplerian orbital elements of the WASP-47 planets.  Combining RVs and TTVs provides a better estimate of the mass of planet d (\md~\mearth) than obtained with only RVs ($12.75\pm2.70~\mearth$) or TTVs ($16.1\pm3.8~\mearth$).  Planets e and d have high densities for their size, consistent with a history of photo-evaporation and/or formation in a volatile-poor environment.   Through our RV and TTV analysis, we find that the planetary orbits have eccentricities similar to the solar system planets.  The WASP-47 system has three similarities to our own solar system: (1) the planetary orbits are nearly circular and coplanar, (2) the planets are not trapped in mean motion resonances, and (3) the planets have diverse compositions.  None of the current single-process exoplanet formation theories adequately reproduce these three characteristics of the WASP-47 system (or our solar system).  We propose that WASP-47, like the solar system, formed in two stages: first, the giant planets formed in a gas-rich disk and migrated to their present locations, and second, the high-density sub-Neptunes formed in situ in a gas-poor environment.
\end{abstract}

\section{Introduction}
One of the key questions driving exoplanet science is the formation of planetary systems in general and the solar system in particular.  The \textit{Kepler} Mission \citep{Borucki2010,Koch2010} led to a wealth of statistical measurements that provide valuable insight into planet formation.  Small planets close to their stars are a common outcome of planet formation \citep{Howard2012, Fang2012, Batalha2013, Fressin2013, Petigura2013_jan, Dressing2015Mdwarfs}, such that half of sun-like stars have at least one planet smaller than Neptune within the orbital distance of Mercury \citep{Petigura2013,Burke2015}.  Many small planets close to their stars are in multi-planet systems \citep{Lissauer2012, Fabrycky2014, Rowe2014, Lissauer2014}.  Given the abundance of small, compact planetary systems around other stars, is our own solar system unusual in that it is barren from Mercury's orbit inward?  Because the \textit{Kepler} Mission only obtained 4 years of continuous photometry and only observed 150,000 stars, it had poor sensitivity to long-period planets, which are unlikely to transit.  If \textit{Kepler} were pointed at our solar system and were lucky enough to discover the inner planets, it still most likely would have missed the planets from Mars out.  How can we reconcile planet formation theory for the close-in exoplanets with planet formation theory for our spaciously spread solar system?

WASP-47 is a system of unusual architecture that might be a Rosetta Stone for linking the exoplanet population to the solar system.  WASP-47 contains a transiting, Jupiter-size planet with an orbital period of 4.2 days (a ``hot Jupiter'') that was detected from the ground-based WASP-South transit survey \citep[WASP-47 b]{Hellier2012}.  What makes WASP-47 b unusual is that, contrary to the vast majority of hot Jupiters, which do not have nearby planetary companions \citep{Steffen2012,Bryan2016}, WASP-47 b has two nearby neighbors: an interior, transiting planet with an orbital period of less than a day (WASP-47 e) and an exterior, transiting planet with an orbital period of 9.0 days (WASP-47 d).  The system also has a distant, moderately eccentric planet (WASP-47 c).  While the compactness of the WASP-47 inner planetary system is comparable to other Kepler systems, especially those that contain ultra-short period planets \citep{Sanchis-Ojeda2013}, the combination of the compact planetary system with a hot Jupiter is unprecedented among the 2217\footnote[1]{Based on a 2016-11-24 query of Exoplanets.org.} planetary systems studied to date.  The architecture of WASP-47 was not predicted by planet formation theory, and so uncovering a physically plausible formation mechanism for WASP-47 will deepen our understanding of planet formation in general.

To better understand which of the various physical models of planet formation and evolution were important in the WASP-47 system, we would  like to measure the masses, densities, bulk compositions, and orbital dynamics of all the planets as precisely as the current data permit.  The compositions of the planets might provide clues about where they formed within the proto-planetary disk--for instance, if the planets are rich in water or other high mean-molecular weight volatiles, they might have formed beyond a molecular snowline.  Furthermore, the present-day orbital elements for the planets can be related to their dynamical history: the semi-major axes and eccentricities of the planets today relate to how they have exchanged energy and angular momentum in the past.

Several analyses of this system have already characterized various dynamical properties of the WASP-47 planets.  The discovery paper \citep{Hellier2012} used two years of ground-based photometry to find WASP-47 b in transit, and also obtained 19 low-precision radial velocities (RVs) to measure the planet's mass.  \citet[][hereafter B15]{Becker2015} discovered two additional transiting planets (e and d) in transits from K2, characterized the planet masses with transit timing variations (TTVs), and used the Mercury N-body integrator \citep{Chambers1999} to explore the dynamical stability of the planets.  \citet{Sanchis-Ojeda2015} measured the projected spin-orbit obliquity of the hot Jupiter via the Rossiter-McLaughlin effect, finding that the planetary orbital axis and the stellar spin axis are not strongly misaligned.  \citet{Dai2015} obtained high-cadence precision RVs of the system, precisely characterizing the mass of the giant planet and placing new mass constraints on the other transiting planets.  \citet{VanMalle2016} discovered a long-period giant planet with a multi-year baseline of radial velocities.  \citet{Almenara2016} simultaneously modeled the K2 light curve and the literature RVs, arriving at planet masses that were determined to a precision of $\sim40\%$.  \citet[][hereafter S17]{Sinukoff2017_W47} obtained 47 new RVs with Keck-HIRES, which, when combined with the literature RVs, significantly improved the precision of the mass and \msini\ measurements ($<25\%$) for all the WASP-47 planets.

We present the a robust analysis of the planet masses and orbital dynamics by combining the 108 transit times measured in B15 with 118 literature radial velocities.  Our paper is structured as follows: in Section \ref{sec:measurements} we introduce the measurements analyzed herein, in Section \ref{sec:TTVs_only} we present two ways to analyze the TTVs alone: using an N-body integrator and a dynamical analytic solver.  In Section \ref{sec:TTV+RV} we present a joint analysis to the TTVs and RVs of the WASP-47 system that results in the most accurate and precise dynamical parameters to date.  In Section \ref{sec:multiply} we present a new, simple way to combine information from TTVs and RVs.  In Section \ref{sec:discussion} we discuss how our improved mass and eccentricity information relates to planet formation theory.  We conclude in Section \ref{sec:conclusion}.

\section{Measurements}
\label{sec:measurements}
The measurements we use in this analysis are all available in the literature.  We combine the 108 transit times (TTs or TTVs) of WASP-47 e, b, and d (B15) with 118 measurements of the radial velocity (RV) of the WASP-47 host star from \citet{Hellier2012}, \citet{Dai2015}, \citet{VanMalle2016}, and S17.  

To combine the RV measurements, we use the values for the RV zero-point offset ($\gamma$) and jitter ($\sigma_\mathrm{jit}$) determined in S17. The zero-point offset is added to each RV measurement, and the jitter is added to each RV uncertainty in quadrature.  
For the \citet{Hellier2012} CORALIE RVs, these values are $\gamma=−27070.3~\ms$, $\sigma_\mathrm{jit} = 5.9~\ms$.
For the \citet{VanMalle2016} CORALIE RVs, these values are $\gamma=−27085.3~\ms$, $\sigma_\mathrm{jit} = 6.7~\ms$.
For the \citet{Dai2015} Magellan-PFS RVs, these values are $\gamma=20.5~\ms$, $\sigma_\mathrm{jit} = 6.3~\ms$. 
For the S17 Keck-HIRES RVs, these values are $\gamma=6.4~\ms$, $\sigma_\mathrm{jit}=3.7~\ms$.
For simplicity, we keep the values of the zero-point offset and jitters fixed at the values determined in S17.  The zero-point offsets and jitter are statistical properties of the RVs, and so we do not expect the TTVs to provide any new information about these parameters.

\section{TTVs-Only Analysis with N-body and Analytic Approaches}
\label{sec:TTVs_only}
In this section, we fit the WASP-47 TTVs as measured in B15 using two different approaches.  First, we do a full N-body simulation of the three transiting planets to reproduce the observed TTVs using the publicly available code \texttt{TTVFast} \citep{Deck2014}.  Then, we use \texttt{TTVFaster} \citep{Agol2016, Agol2016TTVFaster}, a publicly available code that analytically models orbits to first order in eccentricity.  The \texttt{TTVFaster} code was designed to reproduce both low-frequency sinusoidal features and high-frequency ``chopping'' patterns in the TTVs.  In the tests below, we determine that \texttt{TTVFaster}, which is orders of magnitude faster than \texttt{TTVFast}, is appropriate for modeling the TTVs in the WASP-47 system.

\subsection{Modeling transit times with an N-body Integrator}
\begin{deluxetable}{cc}
\tabletypesize{\scriptsize}
\tablecaption{Priors on Dynamical Parameters}
\tablehead{\colhead{Parameter} & \colhead{Priors}} 
\startdata
$M$ & $M$ > 0, Hill criterion   \\
$P$ & $P$ > 0, Hill criterion \\
$TT$ & None \\
$\sqrt{e}$cos$\omega$ & $e < 0.06$, Hill criterion  \\
$\sqrt{e}$sin$\omega$ & $e < 0.06$, Hill criterion  \\
$\sqrt{e_c}$cos$\omega_c$ & $e < 1$, Hill criterion  \\
$\sqrt{e_c}$sin$\omega_c$ & $e < 1$, Hill criterion  \\
\enddata
\label{tab:priors}
\end{deluxetable}
We used the publicly available N-body integrator \texttt{TTVFast} \citep{Deck2014} with the python wrapper \texttt{ttvfast-python}\footnote{https://github.com/mindriot101/ttvfast-python} to forward-model the transit times of the inner three planets.  Unlike in the B15 analysis, we allowed all of the initial osculating elements, particularly the orbital periods and initial times of transits, to vary.  Thus, the variables for each planet $k$ are: the mass of the planet $M_k$, the orbital period $P_k$, the first time of transit $tt_k$, and the eccentricity parametrization $\sqrt{e_k}$cos$\omega_k$, $\sqrt{e_k}$sin$\omega_k$.  We limited $e < 0.06$ for the three inner planets, in accordance with the 10 Myr stability analysis in B15.  For each adjacent pair of planets, we satisfied the Hill criteria for stability for low-eccentricity orbits, as described in Equations 24 and 28 of \citet{Gladman1993}:

\begin{equation}
p - q > 2.4 (\mu_1^2 + \mu_2^2)^{1/3}
\label{eqn:hill_1}
\end{equation}
\begin{equation}
p - q > \sqrt{8/3 (e_1^2 +e_2^2) + 9 \times \mathrm{max}(\mu_1,\mu_2)^{2/3}}
\label{eqn:hill_2}
\end{equation}
where the subscript 1 refers to the inner planet and 2 refers to the outer planet, $q$ is the apoapse distance of the inner planet $q = 1+e_1$, $p$ is the periapse distance of the outer planet $p = (1-e_2)\frac{a_2}{a_1}$, $e$ is the eccentricity, and $\mu$ is the planet-to-star mass ratio.  We also required the planet masses and orbital periods to have positive values.  Because the orbits of the planets are very nearly coplanar \citep{Becker2015, Almenara2016}, we fixed the orbital inclination and longitude of ascending node in a manner consistent with coplanar, edge-on orbits.  See Table \ref{tab:priors} for a summary of the priors and constraints.
\begin{deluxetable}{lcc}
\tablewidth{\columnwidth}
\tabletypesize{\scriptsize}
\tablecaption{Dynamical Parameters from Best N-Body Fit to TTVs Only (TTVFast)}
\tablehead{\colhead{Parameter} & \colhead{Median$\pm$Std. Dev.} & \colhead{Units}} 
\startdata
$M_e$ & 176 $\pm$ 118  & \mearth\\
$M_b$ & 549 $\pm$ 252& \mearth \\
$M_d$ & 16.1$\pm$3.8 & \mearth\\
$P_e$ & 0.78964$\pm$0.00002  & days \\
$P_b$ & 4.150$\pm$0.006 & days \\
$P_d$ & 9.12$\pm$0.05 & days \\
$TT_e$ & 2146.7639$\pm$0.0008 & BJD - 2454833 \\
$TT_b$ & 2149.969$\pm$0.006  & BJD - 2454833 \\
$TT_d$ & 2155.40$\pm$0.05 & BJD - 2454833 \\
$\sqrt{e_e}$cos$\omega_e$ & -0.006$\pm$0.14 \\
$\sqrt{e_e}$sin$\omega_e$ & -0.008$\pm$0.14 \\
$\sqrt{e_b}$cos$\omega_b$ & 0.001$\pm$0.08 \\
$\sqrt{e_b}$sin$\omega_b$ & 0.006$\pm$0.07 \\
$\sqrt{e_d}$cos$\omega_d$ & -0.09$\pm$0.11 \\
$\sqrt{e_d}$sin$\omega_d$ & 0.07$\pm$0.09 \\
\enddata
\tablecomments{These are the initial astrocentric Keplerian orbital elements, reported at epoch BJD 2456979.4961.  They are not the time-averaged orbital properties of the planets.}
\label{tab:n-body}
\end{deluxetable}

\begin{figure}
    \centering
    \includegraphics[width=\columnwidth]{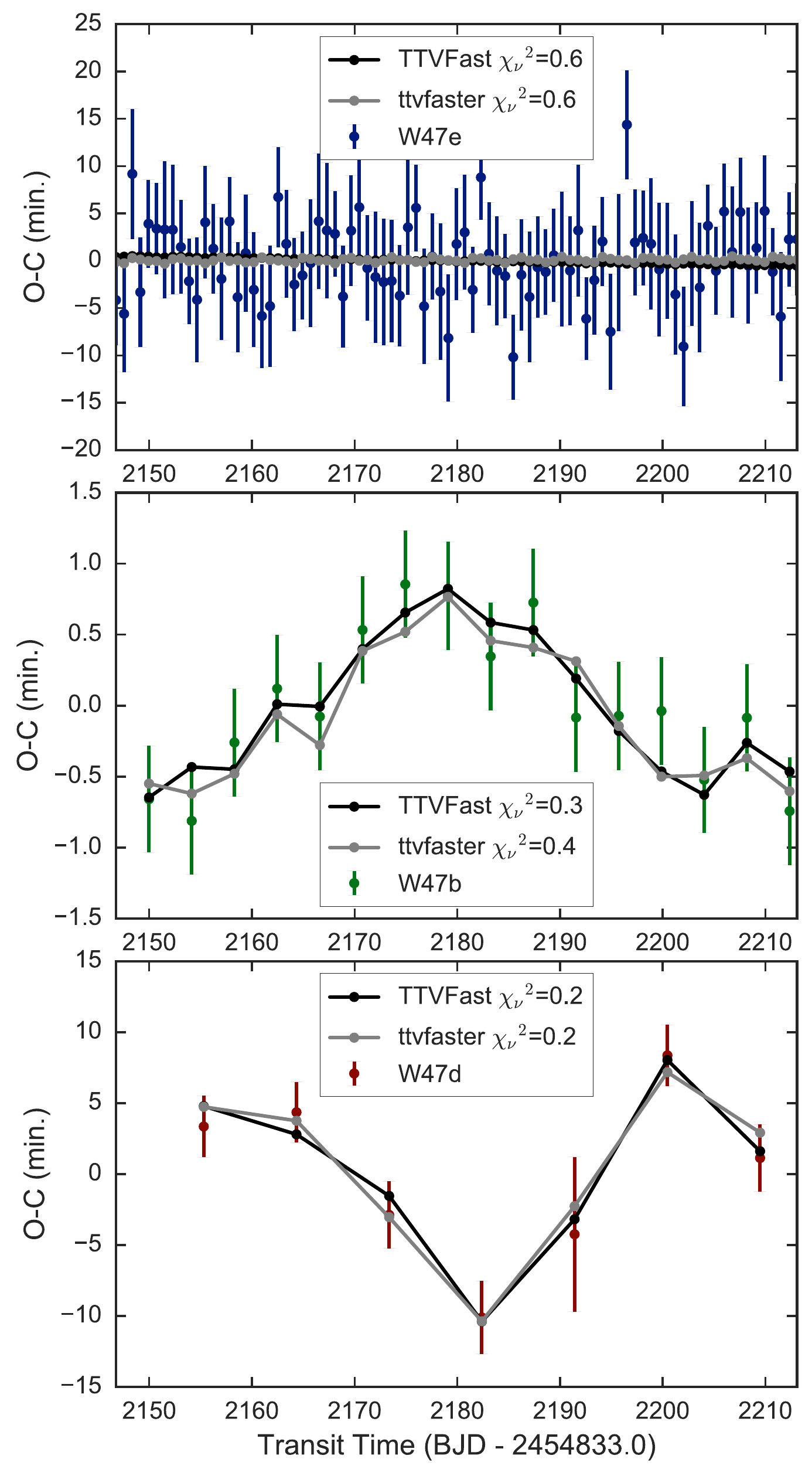}
    \caption{The observed minus linear-ephemeris calculated transit times of WASP-47 e (blue points, top panel), b (green points, middle panel), and d (brown points, bottom panel).  The best-fit N-body model to the TTVs alone (using \texttt{TTVFast}, black connected dots) and the best-fit analytic model to the TTVs alone (using \texttt{TTVFaster}, gray connected dots) are shown.}
    \label{fig:ttv-compare}
\end{figure}

\begin{figure*}
    \centering
    \includegraphics[width=2\columnwidth]{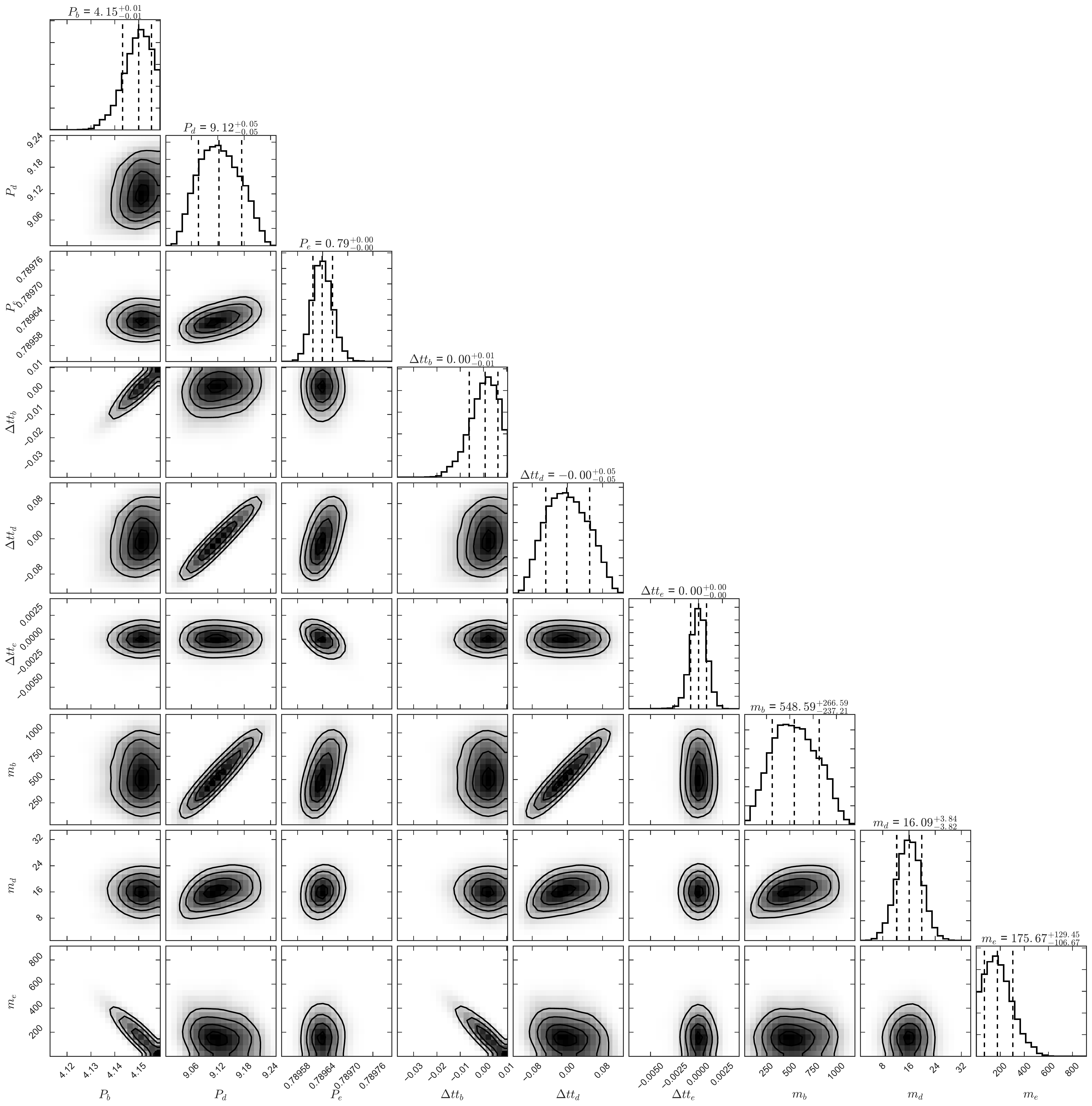}
    \caption{Posteriors of the planet masses, initial osculating orbital periods, and initial times of transit for WASP-47 e, b, and d in the \texttt{TTVFast} (N-body) analysis.  The planet masses are highly covariant with the initial orbital elements.  Note that the orbital periods and times of transit here are initial osculating elements, not the time-averaged orbital elements.  The MCMC chains shown are available as the Data behind the Figure.}
    \label{fig:nbody-post}
\end{figure*}

We used the Markov Chain Monte Carlo (MCMC) Python package \texttt{emcee} \citep{Foreman-Mackey2013} to explore the posteriors of various combinations of the dynamical parameters.  We ran 60 walkers $5\times10^5$ steps each, throwing away the first $10^5$ steps as burn-in, and checked that the multivariate extension of the potential scale reduction factor (PSRF) statistic \citep[$\hat{R} < $1.2,][]{Gelman1992, Brooks1998} converged.  We also inspected the chains by eye to check for convergence.  Our best fit\footnote{from the MCMC maximum likelihood} to the transit times using \texttt{TTVFast} is shown in Figure \ref{fig:ttv-compare}.  Table \ref{tab:n-body} summarizes our results from this N-body fit to the TTVs.

We find that, for planets e and b, the masses and eccentricities of the planets are highly covariant with the \textit{initial osculating orbital period} and the \textit{initial transit time} of neighboring planets (see Figure \ref{fig:nbody-post}).  This is because the initial osculating orbital period is translated to an instantaneous velocity and acceleration, and the planet's acceleration depends on the mass and position of its neighbor.  Therefore, it is critical to allow the initial orbital periods and times of transit of all the planets to vary in order to explore the full range of possible planet masses.  Furthermore, only one super-period of the TTVs is observed, and so an average orbital period and a representative time of transit are not as well determined for planets b and d as they might be with the observation of multiple super-periods. 
Thus, we find that the TTVs do not constrain the masses of WASP-47 e or WASP-47 b as tightly as what is reported in B15.  Whereas B15 find $M_b=341^{+73}_{-55}~\mearth$, we find $M_b=549\pm252~\mearth$, using the exact same TTV measurements.  The mass for planet e reported in B15 stems from their choice of prior: they find $M_e < 22~\mearth$, whereas we find $M_e=176\pm118~\mearth$.  However, the TTVs of planet b place strong constraints on the mass of planet d.  B15 find $M_d = 15.2\pm7~\mearth$, and we find $M_d = 16.1\pm3.8~\mearth$.   While an analytic covariance between planet masses and the free eccentricity exists \citep{Lithwick2012}, our stability constraint that $e < 0.06$ minimizes the effects of this degeneracy.

\subsection{Modeling transit times analytically}
We used the publicly available analytic TTV package \texttt{TTVFaster} to model the transit times of the inner three planets observed in B15.  The \texttt{TTVFaster} code analytically transforms the planet mass $M_k$ and average orbital elements $P_k$, $tt_k$, $e_k$cos$\omega_k$, $e_k$sin$\omega_k$\footnote{We used jump parameters $\sqrt{e_k}\mathrm{cos}\omega_k$, $\sqrt{e_k}\mathrm{sin}\omega_k$ to avoid an eccentricity bias and speed convergence.} into a TTV pattern, to first order in eccentricity, to a user-specified precision in the disturbing function.  We found that modeling to sixth order in the expansion of the Laplace coefficient $b_{1/2}^j$ ($j = \{0,1,2,3,4,5,6\}$), \citep{Murray&DermottCh8} was sufficient to reproduce the observed TTV signature with the same fidelity as produced in the N-body analysis (see Figure \ref{fig:ttv-compare}).  In general, \texttt{TTVFaster} is designed to work for planets that (1) are not extremely close to a mean motion resonance, (2) have low eccentricities, (3) have low masses.  Because WASP-47 b is a Jupiter-mass planet, we wanted to see if \texttt{TTVFaster} modeled the orbital dynamics correctly.

Incorporating the priors from Table \ref{tab:priors}, we used Python packages \texttt{lmfit} \citep{Newville2014} and \texttt{emcee} to explore the posteriors of the dynamical parameters.  We ran 100 walkers 20,000 steps, throwing away the first 4000 steps as burn-in.  We note that the chains converged much faster (according to the PSRF statistic) when we used \texttt{TTVFaster} than when we used \texttt{TTVFast}, because the TTVs provide better constraints on the average orbital parameters than they do on the initial orbital parameters.  

\begin{figure*}
    \centering
    \includegraphics[width=2\columnwidth]{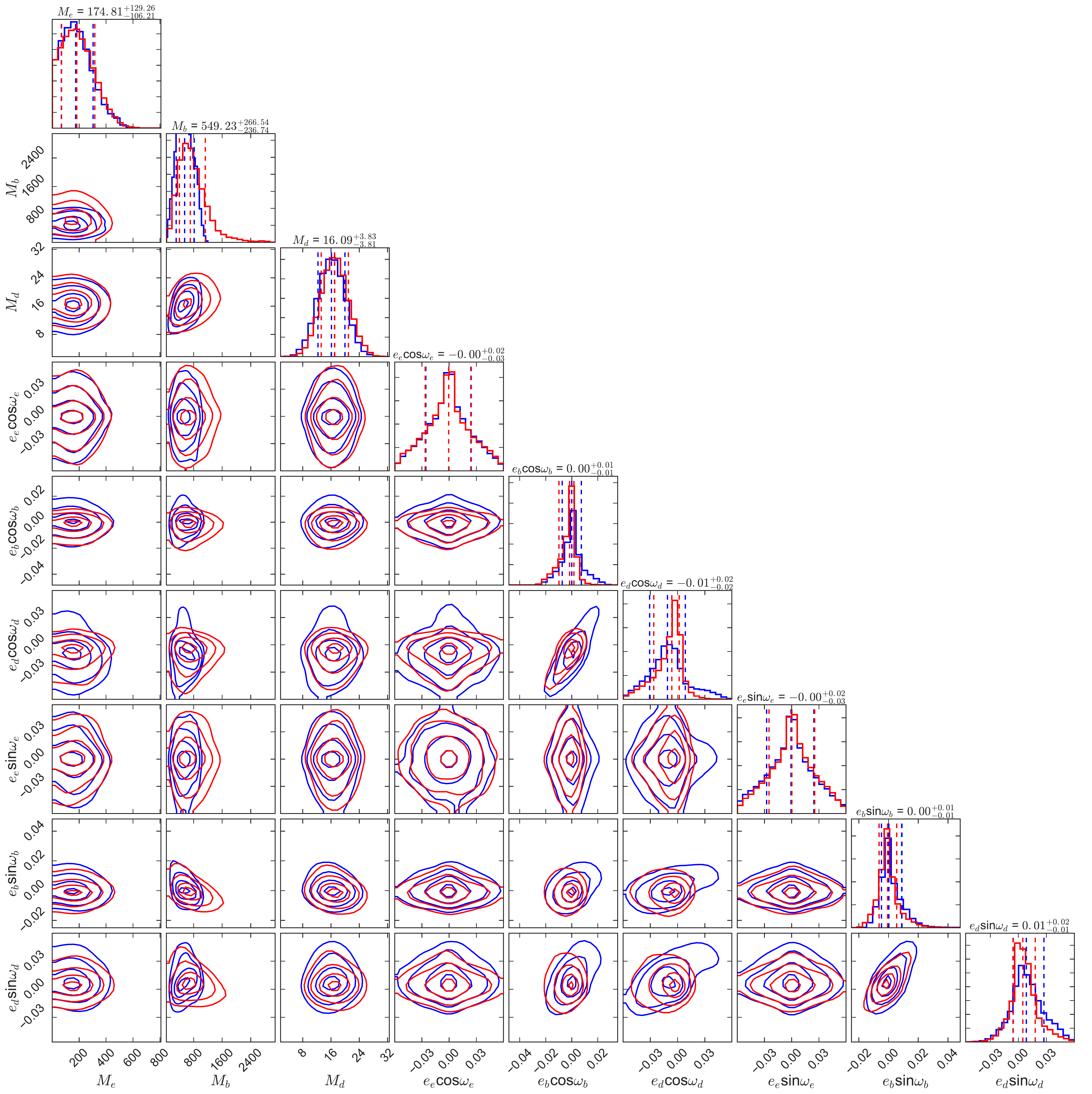}
    \caption{The posteriors of the analyses of the TTVs alone.  Blue: from the \texttt{TTVFast} N-body integrator MCMC; red: from the analytic TTV modeler TTVFaster MCMC.  In both analyses, we limited $e < 0.06$ for the three inner planets, in accordance with the stability analysis from B15.  We also required Hill stability.  We used jump parameters of the form $\sqrt{e}$cos$\omega$ to avoid high eccentricity biases.  The mass and eccentricity posteriors obtained from the analytic TTV analysis are in good agreement with those obtained with N-body modeling.  The values above the histograms correspond to the N-body posterior median and $1\sigma$ bounds.  The MCMC chains shown are available as the Data behind the Figure}
    \label{fig:posterior_compare}
\end{figure*}

The mass and eccentricity distributions we determined from the analytic solution to the observed TTVs are consistent with the N-body model (Figure \ref{fig:posterior_compare}).  Since fitting an analytic model to the TTVs is orders of magnitude faster than a full N-body analysis (especially when the long time baseline for RVs is required), we use the analytic modeling technique in the rest of this paper.

\section{Combining TTVs and RVs with \texttt{TTVFaster}}
\label{sec:TTV+RV}
We combined the python packages \texttt{TTVFaster} and \texttt{radvel} (Fulton \& Petigura in prep.\footnote{https://github.com/California-Planet-Search/radvel}) to simultaneously fit the RVs and TTVs, resulting in refined masses and orbital properties of all four known planets.  To fit the RV and TTV data simultaneously, we maximized the following log-likelihood function while satisfying our priors:
\begin{equation}
\label{eqn:logL}
\begin{split}
\mathrm{ln}L = - \sum_i^{N_\mathrm{RV}} \frac{(RV_{\mathrm{obs},i}- RV_{\mathrm{Kep},i})^2}{2\sigma_{\mathrm{RV},i}^2} \\
- \sum_{k}^{N_\mathrm{pl}}\sum_{j}^{N_\mathrm{TT,k}} \frac{(TT_{\mathrm{obs},k,j} - TT_{\mathrm{model},k,j})^2}{2\sigma_\mathrm{TT,k,j}^2}\\
\end{split}
\end{equation}
where $N_\mathrm{RV}$ is the number of RVs; $N_\mathrm{pl}$ is the number of planets; $N_{\mathrm{TT},k}$ is the number of transit times for planet $k$; $RV_{\mathrm{obs},i}$ is the $i$th observed RV, including the instrument-specific fixed zero-point offset $\gamma$ determined in S17; $RV_{\mathrm{Kep},i}$ is the $i$th Keplerian-modeled RV; $\sigma_{\mathrm{RV},i}$ is the uncertainty in $RV_{\mathrm{obs},i}$, including a constant jitter for each spectrometer determined in S17; $TT_{\mathrm{obs},k,j}$ is the $j$th observed transit time for planet $k$; $TT_{\mathrm{model},k,j}$ is the $j$th modeled transit time for planet $k$; and $\sigma_{\mathrm{TT},k,j}$ is the uncertainty in $TT_{\mathrm{obs},k,j}$.

Our model included all four known planets.  The variable parameters for each planet $k$ are: the mass of the planet $M_k$, the orbital period of the planet $P_k$, a representative time of transit $tt_k$, and the eccentricity parametrization $\sqrt{e_k}$cos$\omega_k$, $\sqrt{e_k}$sin$\omega_k$.  Note that the argument of periapse passage, $\omega_k$, is for the planet, not the star.  This formulation is consistent with both the \texttt{TTVfaster} definition and the definition in \citet{Seager_exoplanets2010_murray} and \citet{Seager_exoplanets2010_lovis}.

\begin{figure}
    \centering
    \includegraphics[width=\columnwidth]{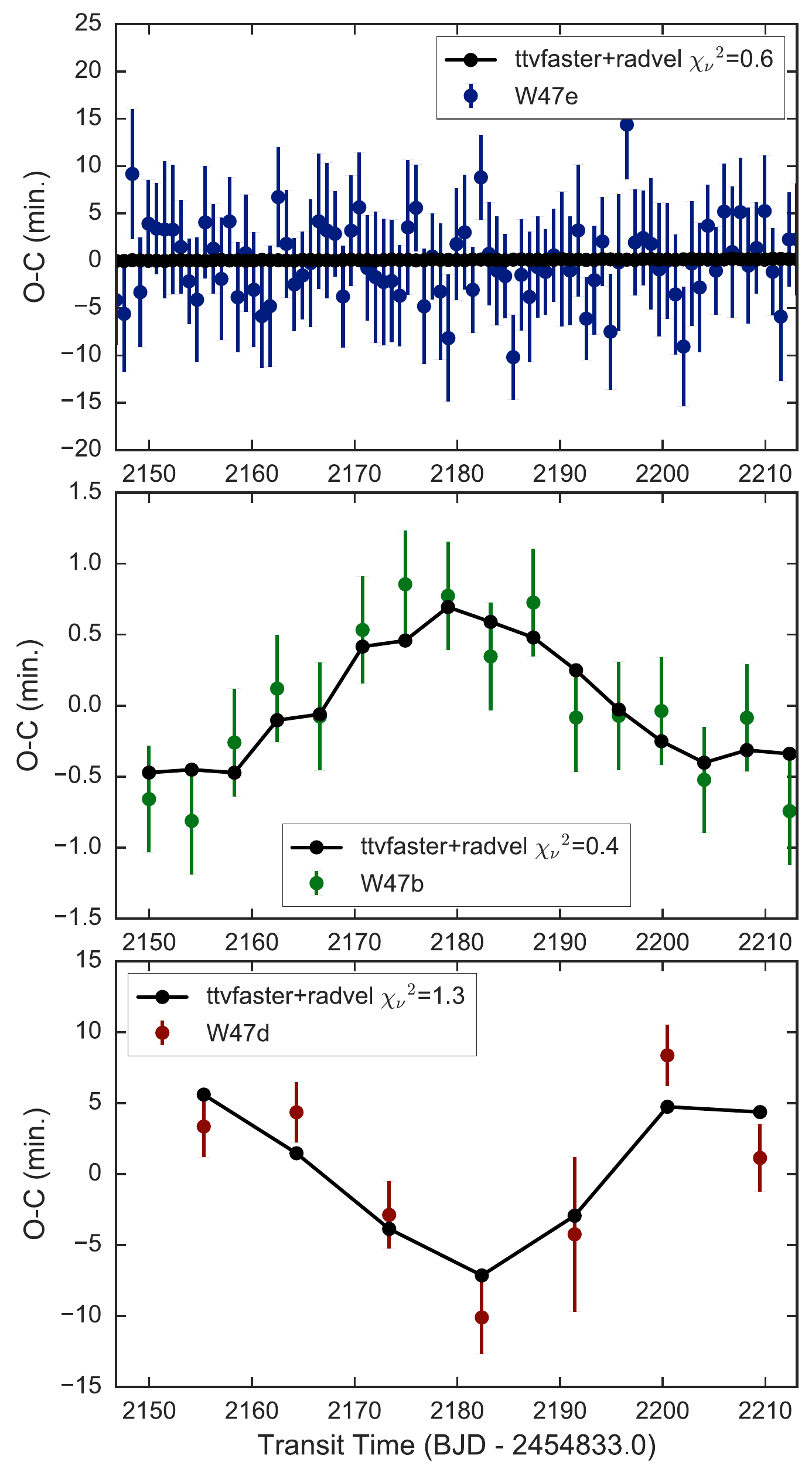}
    \caption{The best simultaneous fit to the TTVs and RVs of the WASP-47 system, and residuals.  From top to bottom, the panels show the TTVs of the transiting planets e ($P=0.79$ days, blue), b ($P=4.16$ days, green), and d ($P=9.0$ days, brown).  The best simultaneous fit to the TTVs + RVs of all four planets is shown as black connected points.}
    \label{fig:TTVFaster+radvel-ttvs}
\end{figure}

\begin{figure*}
    \centering
    \includegraphics[width=\columnwidth]{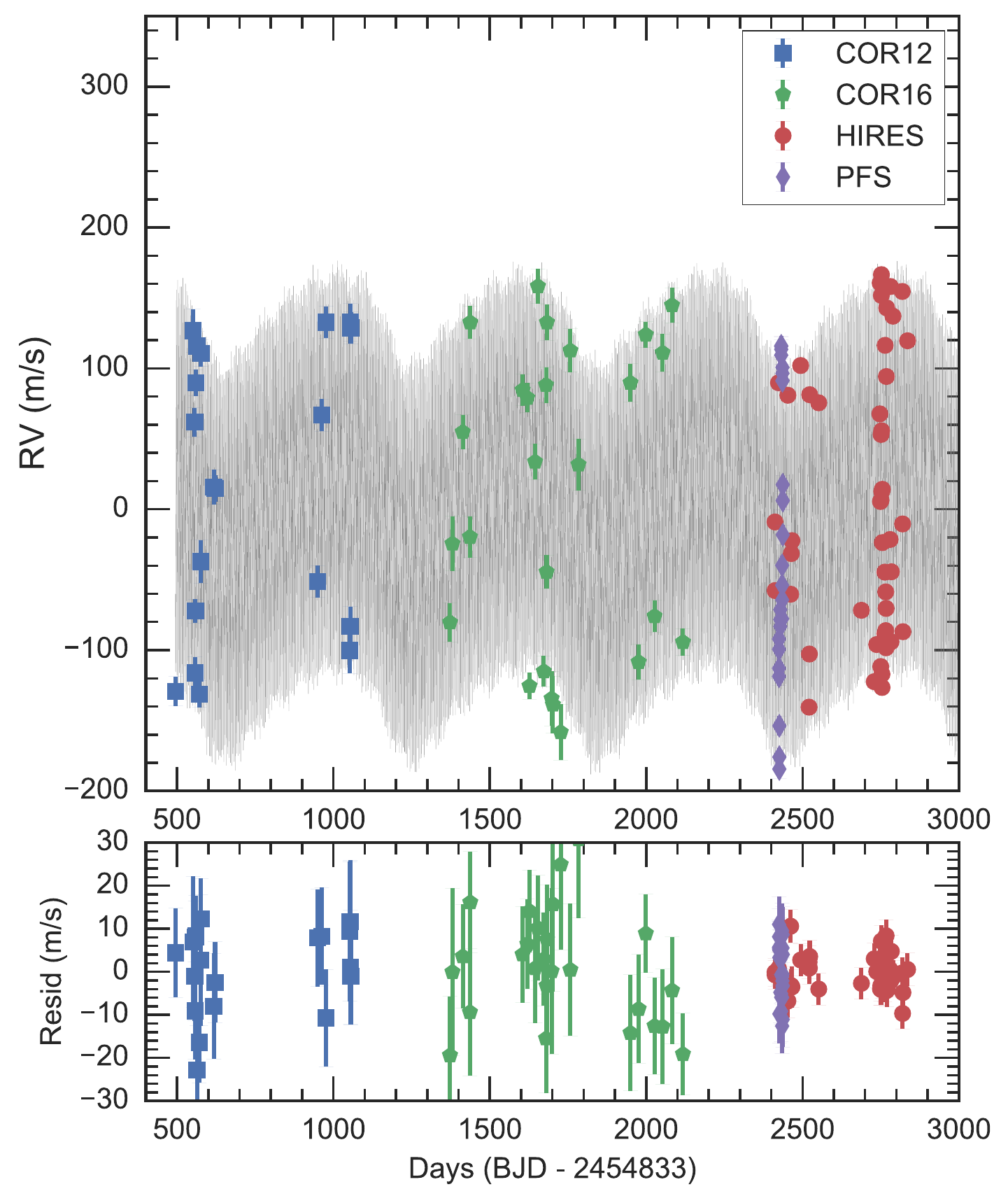}
    \includegraphics[width=\columnwidth]{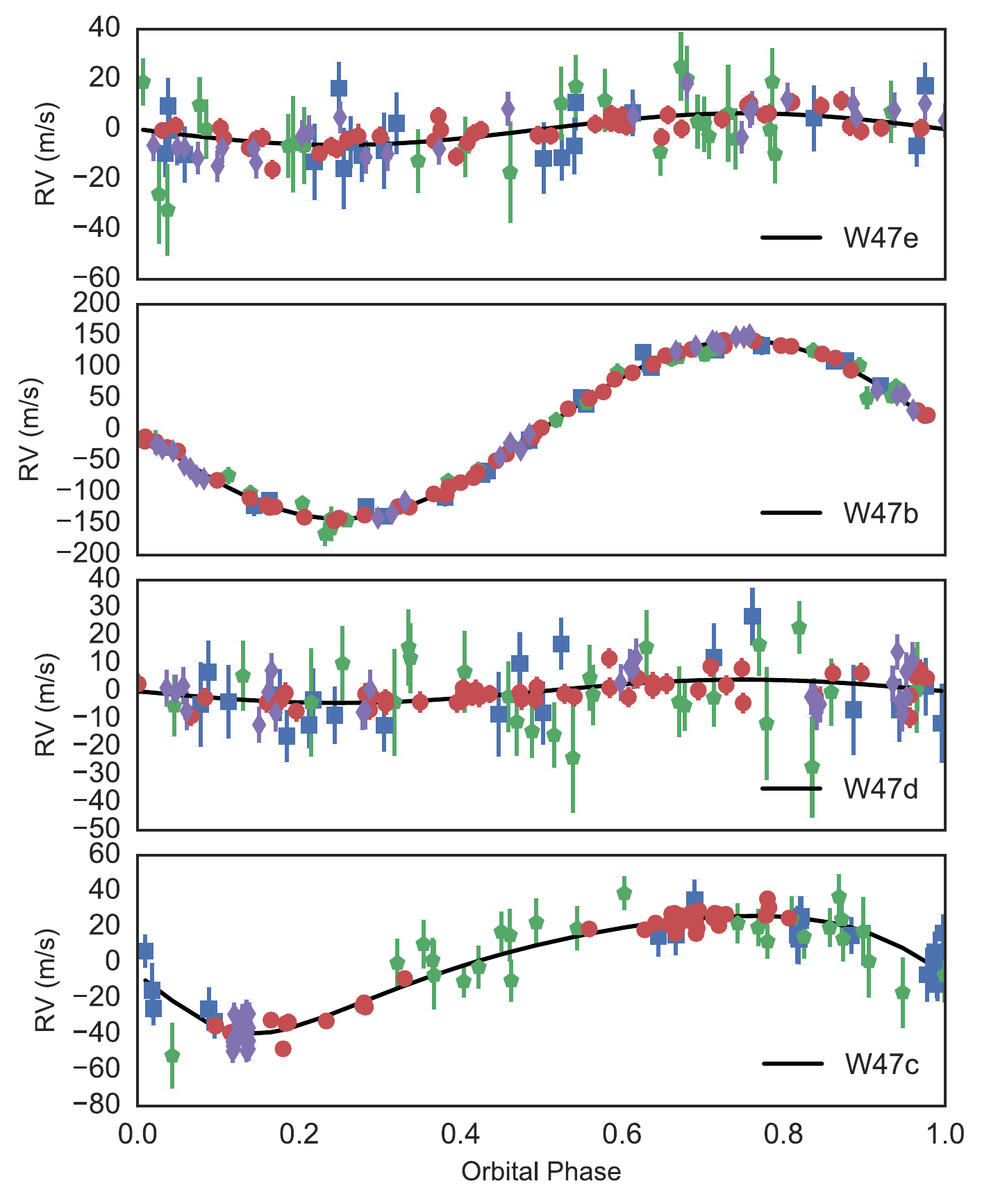}
    \caption{Left, top: Radial velocities of WASP-47 from four observational campaigns: CORALIE before 2012 (blue squares), CORALIE before 2016 (green pentagons), PSF (purple diamonds), and HIRES (red circles).  The best-fit model to the TTVs and RVs (fine gray line) is shown.  Left, bottom: RV residuals (observations minus the \texttt{TTVFaster+radvel} model values).  The RMS of the residuals is 8.5 \ms, which is comparable to the mean jitter-enhanced RV uncertainty over all the telescopes (7.7 \ms).  Right: the RVs phase-folded to the orbital periods of planet e (top), b (second from top), d (second from bottom), and c (bottom).  The HIRES RVs are the only single dataset that constrain the semi-amplitudes of all the planets, because they have the precision (3 \ms) to capture the small amplitudes of planets e and d, and also the baseline to capture the amplitude of the long-period planet c.}
    \label{fig:TTVFaster+radvel-rvs}
\end{figure*}

These parameters are transformed into the appropriate basis to drive a Keplerian RV model (for comparison to the RVs) and the basis used for \texttt{TTVFaster} computations.  Note that this scheme is not possible for an N-body integrator, since the initial orbital elements are not the same as the time-averaged orbital elements used in a Keplerian prescription.  We also required Hill stability for all the planets, as described in Equations \ref{eqn:hill_1} and \ref{eqn:hill_2}.  In addition, we allowed the stellar mass to vary, using the prior $\mstar = 0.99\pm0.05~\msun$ from S17, in case the combined RV and TTV data added new information about the stellar mass.\footnote{The TTVs constrain $M_k/\mstar$, whereas the RVs constrain $M_k/\mstar^{2/3}$.}  The best simultaneous fit to the TTVs and RVs is shown in Figures \ref{fig:TTVFaster+radvel-ttvs} (TTVs) and \ref{fig:TTVFaster+radvel-rvs} (RVs).

\floattable
\begin{deluxetable}{lccccr}
\tabletypesize{\scriptsize}
\tablecaption{Dynamical Parameters from Simultaneous Fit to TTVs (ttvfaster) + RVs (Keplerian)}
\tablehead{\colhead{Parameter} & \colhead{Median $\pm$ Std. Dev.} & \colhead{95\% U.L.} & \colhead{Units} & \colhead{Ref.}} 
\startdata
\multicolumn{4}{l}{\textit{MCMC jump parameters}}\\
\mstar & \stellarmass & & \msun & A\\
$M_e$ & \me && \mearth & A\\
$M_b$ & \mb && \mearth & A\\
$M_d$ & \md && \mearth & A\\
$M_c\mathrm{sin}i_c$ & \mcsini && \mearth & A\\
$P_e$ &\pere && days & A\\
$P_b$ & \perb && days  & A\\
$P_d$ & \perd && days & A\\
$P_c$ & \perc && days & A\\
$TT_e$ & \tte && KJD$^\textbf{b}$ & A\\
$TT_b$ & \ttb && KJD & A\\
$TT_d$ &  \ttd && KJD  & A\\
$TT_c$ &  \ttc && KJD & A\\
$\sqrt{e_e}$cos$\omega_e$ & \secwe && & A\\
$\sqrt{e_b}$cos$\omega_b$ & \secwb &&  & A\\
$\sqrt{e_d}$cos$\omega_d$ & \secwd &&  & A\\
$\sqrt{e_c}$cos$\omega_c$ & \secwc &&  & A\\
$\sqrt{e_e}$sin$\omega_e$ & \seswe && & A\\
$\sqrt{e_b}$sin$\omega_b$ & \seswb  &&  & A\\
$\sqrt{e_d}$sin$\omega_d$ & \seswd  && & A\\
$\sqrt{e_c}$sin$\omega_c$ & \seswc && & A\\
\hline
\multicolumn{4}{l}{\textit{Parameters from photodynamical analysis}} \\
\rhostar &  $0.999\pm0.015$ & & \gcc & B \\
$R_e/\rstar$ & $0.01439\pm0.00016$ & & & B\\
$R_b/\rstar$ & $0.10193\pm0.00018$ & & & B\\
$R_d/\rstar$ & $0.02931\pm0.00015$ & & & B\\
\hline
\multicolumn{4}{l}{\textit{Derived Parameters}} \\
\rstar & $1.12\pm0.02$ && \rsun & A,B\\
$R_e$  & \re &&\rearth& A,B \\
$R_b$   &   \rb &&\rearth& A,B\\
$R_d$  &  \rd &&\rearth& A,B\\
$\rho_e$  &    $\rhoe$ &&\gcc& A,B\\
$\rho_b$   &   $\rhob$ &&\gcc& A,B\\
$\rho_d$  &  $\rhod$ &&\gcc& A,B\\
$e_e$ & $0.03\pm0.02$ & $<$\ee$^\textbf{b}$ & & A\\
$e_b$ & $0.0028\pm0.0028$ & $<$\eb & & A\\
$e_d$ & $0.007\pm0.007$ & $<$\ed & & A\\
$e_c$ & \ec & & & A\\
$e_d\mathrm{cos}\omega_d$-$e_b\mathrm{cos}\omega_b$ & $-0.001\pm0.005$ & & & A\\
$e_d\mathrm{sin}\omega_d$-$e_b\mathrm{sin}\omega_b$ & $0.0\pm0.007$ &  & & A\\
$\omega_e$  &    $81.0\pm146.0$ &&deg.& A\\
$\omega_b$   &   $51.0\pm82.0$ &&deg.& A\\
$\omega_d$  &   $76.0\pm106.0$ &&deg.& A\\
$\omega_c$  &  $138.0\pm8.0$ &&deg.& A\\
\enddata
\tablecomments{Results from the MCMC analysis of the TTVs + RVs.  The columns are: parameter, median value plus-or-minus standard deviation, 95\% upper limit (if interesting), and units.  A--Derived in this analysis.  B--incorporating $\rhostar=0.999\pm0.015$ from \citet{Almenara2016}. \textbf{a} KJD = BJD - 2454833.0. \textbf{b}--Note that the upper limit on the eccentricity of planet e is determined from orbital stability requirements, not the measurements.}
\label{tab:params}
\end{deluxetable}

\begin{figure*}
    \centering
    \includegraphics[width=2\columnwidth]{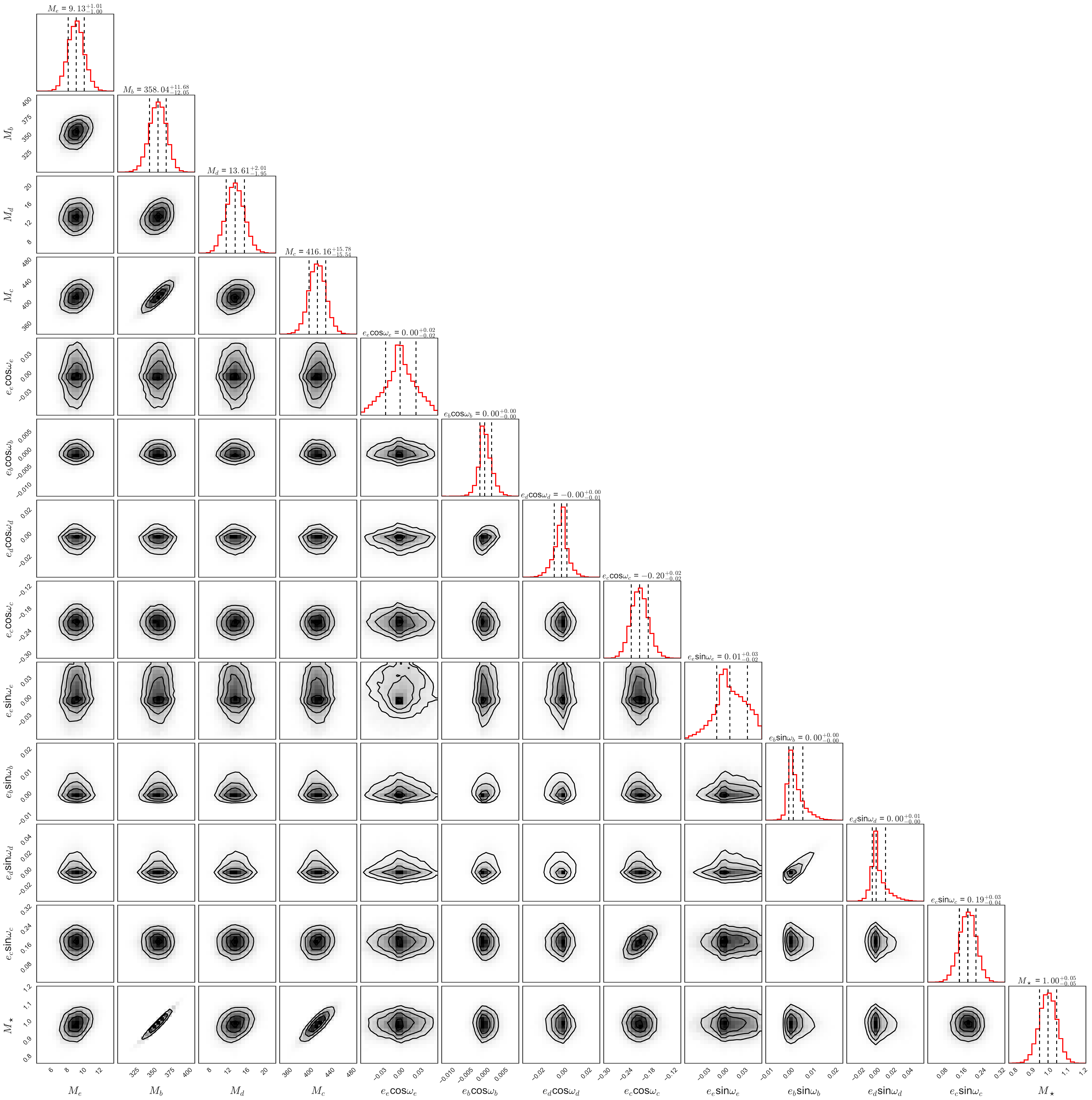}
    \caption{The mass and eccentricity posteriors of the WASP-47 planets based on a simultaneous fit to the TTVs and RVs (see Equation \ref{eqn:logL}) using the analytic TTV package \texttt{TTVFaster} and the Keplerian RV package \texttt{radvel}.  The masses, orbital periods, times of transit, $\sqrt{e}$cos$\omega$, $\sqrt{e}$sin$\omega$, and stellar mass were allowed to vary.  The MCMC chains shown are available as the Data behind the Figure}
    \label{fig:TTVFaster-radvel-posterior}
\end{figure*}
  
Incorporating the priors in Table \ref{tab:priors}, we used \texttt{emcee} to explore the posteriors and covariances of the dynamical parameters.  We ran 100 walkers 10,000 steps each, throwing away the first 4000 steps as burn-in, and found that our chains had converged based on the PSRF statistic.  (The inclusion of RV data helped the chains converge faster.)  The result of our MCMC analysis is shown in Figure \ref{fig:TTVFaster-radvel-posterior}. Our MCMC results are summarized in Table \ref{tab:params}.

To compute planet densities, we utilized the precise stellar density determined by the photodynamical analysis in \citet{Almenara2016}.  Because the transit of the giant planet has very high signal-to-noise, the transit ingress and egress are well-resolved.  The clear ingress and egress and the nearly circular orbit of the giant planet enable a precise characterization of the stellar limb darkening parameters and the planet impact parameter.  Knowledge of these physical quantities allows a precise determination of the stellar density.

We translated the precise stellar density into precise planet densities and radii in the following manner.  For each MCMC trial, we drew a random stellar density from a normal distribution $\mathcal{N}(0.999,0.015)$.  We combined the stellar mass and stellar density of each trial to compute the stellar radius.  For each trial, we also drew a random planet-to-star radius ratio for each planet, using the radius ratios determined in \citet[see Table \ref{tab:params}]{Almenara2016}.  We computed the planet density and radius with the following equations:
\begin{equation}
\rpl = \rprs \rstar
\end{equation}
\begin{equation}
\rhopl = \rhostar \left(\mpms\right) \left(\rprs\right)^{-3}.
\end{equation}
The small uncertainty in the stellar density constrains the planet densities, since the stellar mass and radius (and hence planet mass and radius) are correlated.  Including the stellar density information reduces the uncertainties in the planet densities by $\sim20\%$.  Note that this refinement of the planet densities and radii does not affect the planet masses; the masses are determined directly from TTVs and RVs.  Rather, a detailed study of the stellar properties propagates to our interpretation of the planetary properties.

\section{Independent Multiplied Posteriors (IMPs): A quick and daring way to combine datasets}
\label{sec:multiply}
In this section we offer a sanity check of our combined RV+TTV dynamical solution.  Since the RVs and TTVs are independent observations, the marginalized posteriors from their separate analyses can be multiplied together to estimate the joint probability distribution of a parameter of interest.  This is codified in probability theory as
\begin{equation}
P(A \cap B) = P(A) \times P(B)
\end{equation}
if A and B are independent events.  Assuming that the RV time series is independent of the TTV times series\footnote{This is easier to satisfy than the claim that each RV and each TTV is an independent measurement}, the posteriors of the RV-only analysis and the TTV-only analysis can be multiplied together to estimate their joint posterior.  We call the product of multiplying the posteriors from independent data sets an Independent, Multiplied Posterior (IMP).  The IMP loses some of the information of a simultaneous TTV+RV analysis because the RVs and TTVs are interleaved in time, and because any subtle, slight covariances in the posteriors are not captured in the IMP.  

\begin{figure}
\centering
\includegraphics[width=\columnwidth,trim={10 210 10 210},clip]{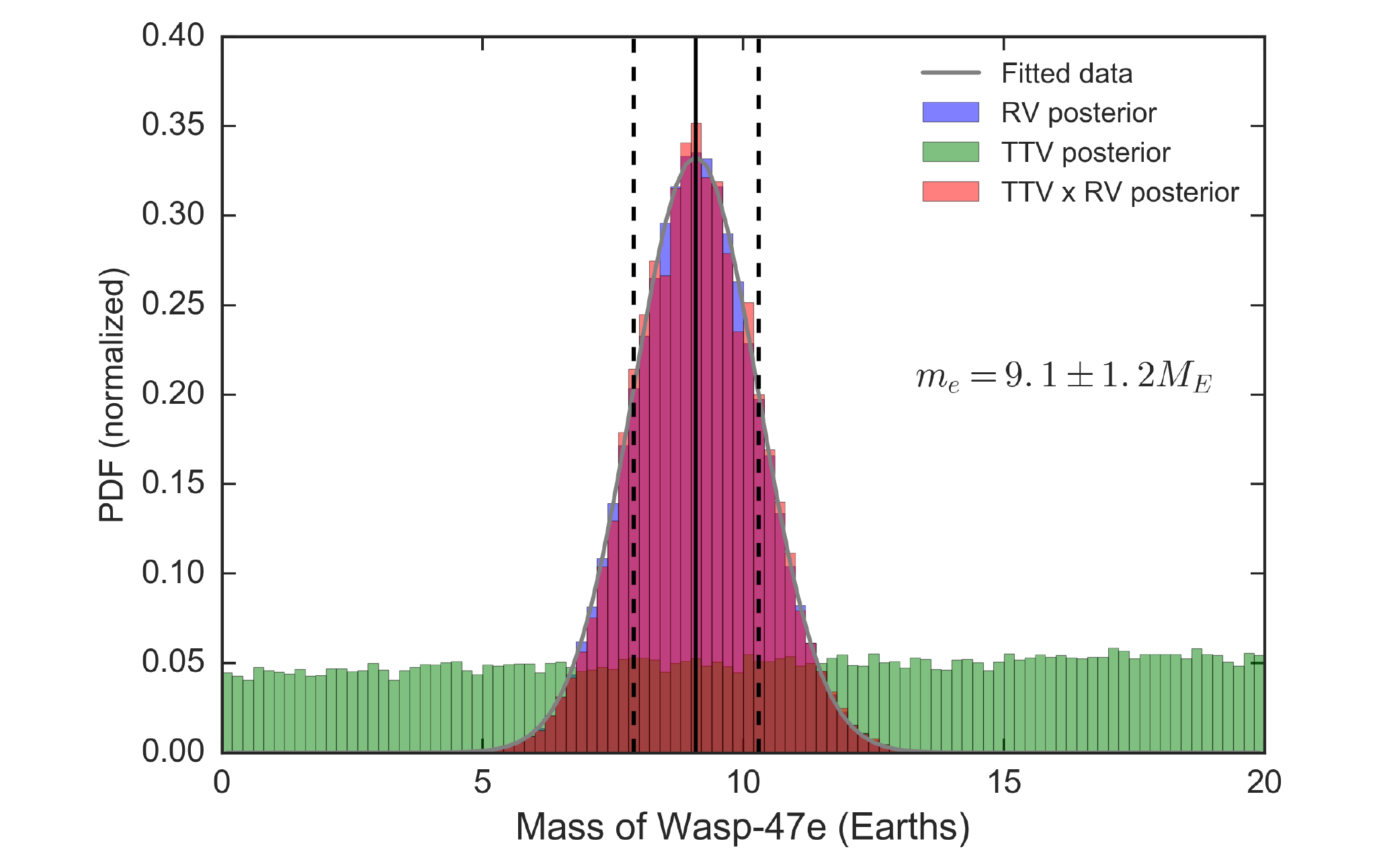}
\includegraphics[width=\columnwidth,trim={10 210 10 210},clip]{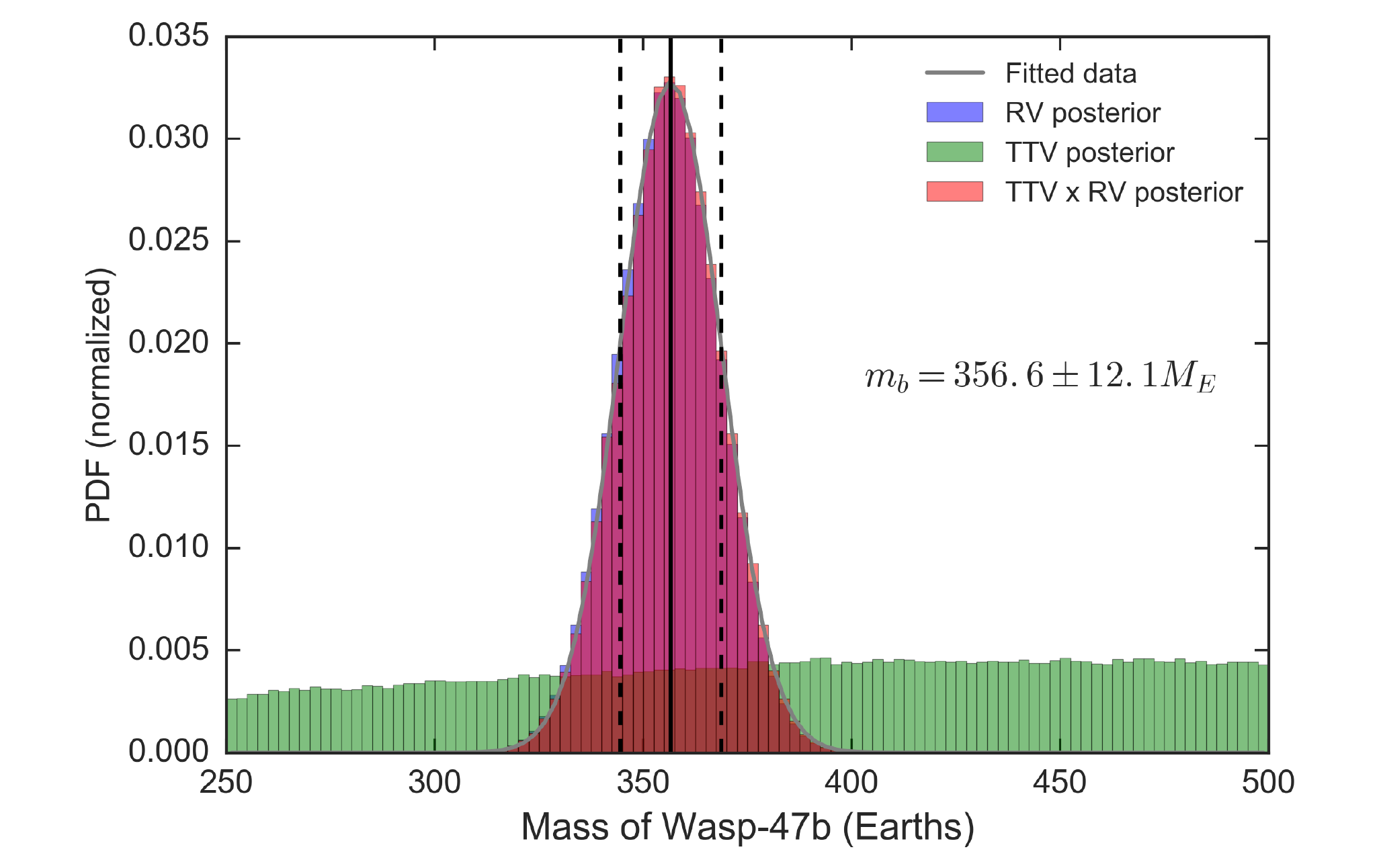}
\includegraphics[width=\columnwidth,trim={10 210 10 210},clip]{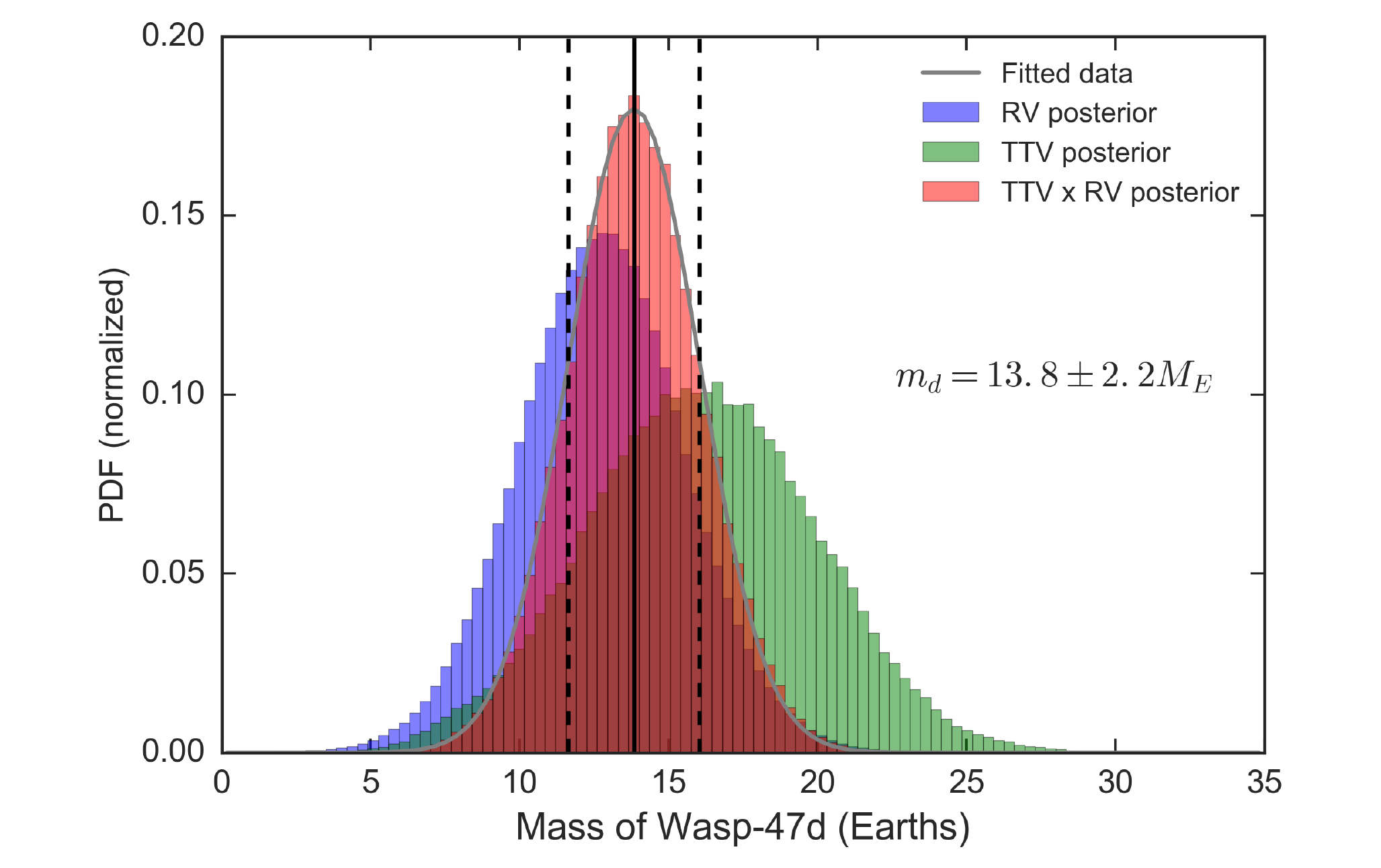}
    \caption{Top: posteriors of the mass of WASP-47 e from analyses of the RVs only (blue), the TTVs only (green, using the N-body integrator TTVFast), and their product (i.e. IMP, red).  The gray line shows a Gaussian fit to the IMP, with the mean (solid black line) and 1$\sigma$ interval (dashed black lines) shown.  Middle: same as the top, but for WASP-47 b.  Bottom: same as the top, but for WASP-47 d.  Note that the result of computing the IMP for planet d is in good agreement with the simultaneous RV+TTV analysis ($\md~\mearth$).}
    \label{fig:imps}
\end{figure}

In Figure \ref{fig:imps}, we show the posteriors of the planet masses from the RV-only (S17) and TTV-only (using the N-body integrator) analyses, and the result of multiplying these posteriors together.  For planets e and b, the TTVs provide no new information, and so the IMPs reflect the mass posteriors from the RV-only analysis.  However, the TTVs alone do measure the mass of planet d.  For planet d, the IMP performs as we would expect: it peaks at a value between the RV-only and TTV-only analyses, and its width is narrower than either analysis is alone.  The IMP for the mass of planet d gives $M_d = 13.8\pm2.2~\mearth$.  This is in good agreement with what we determined in the simultaneous modeling of the RVs+TTVs ($M_d = \md~\mearth$).  As expected, the constraint we get from the simultaneous modeling of the RVs+TTVs is slightly tighter than the constraint from the IMP.  Also, the result of simultaneous modeling is slightly closer to the RVs-only solution ($M_d = 12.75\pm2.70~\mearth$) than the TTVs-only solution ($M_d = 16.1\pm3.8~\mearth$).

In cases where one is computation-limited or short on time, the IMP provides an approximate answer.  However, if the posteriors are highly covariant in both data sets, the IMP might grossly overestimate the uncertainties and might also lose accuracy.  This method for combining data sets is convenient and potentially useful for combining TTV and/or RV data sets with data from GAIA or WFIRST in the future, but should be used with great caution.

\section{Discussion}
\label{sec:discussion}
Here we examine how our joint analysis of the WASP-47 TTVs and RVs provides information about the compositions, orbital dynamics, and formation history of the WASP-47 planets.

\subsection{Relative Information in Dynamical Analyses}
\floattable
\begin{deluxetable}{lcccccc}
\tablecaption{Relative Information in Literature Dynamical Analyses}
\tablehead{\colhead{Parameter} & \colhead{B15} & \colhead{A16.1}& \colhead{A16.2}& \colhead{S16} & \colhead{TTVs-Nbody} & \colhead{RVs+TTVs}} 
\startdata
$\mstar [\msun]$ & $1.04\pm0.08$ & $1.11^{+0.89}_{-0.49}$ & $1.029\pm0.031$ & $0.99\pm0.05$ & $0.99\pm0.05$ & \stellarmass\\
$M_e$ [\mearth]  & $<22^P$ & $9.1^{+5.5}_{-2.9}$ & $9.1^{+1.8}_{-2.9}$ & $9.11\pm1.17$ & $176\pm118$ & \me \\
$M_b$ [\mearth] & $341^{+73}_{-55}$ & $383^{+190}_{-120}$ & $363.8\pm8.6$ & $356\pm12$ & $549\pm252$ & \mb\\
$M_d$ [\mearth] & $15.2\pm7$ & $16.8^{+12}_{-7}$ & $15.7\pm1.1$ &$12.75\pm2.70$ & $16.1\pm3.8$ & \md \\
$M_c$ [\mearth] & -- & $500^{+320}_{-190}$ & $470^{+200}_{-100}$ & $411\pm18$&  --& \mcsini \\
$e_e$ & $< 0.06$ & $< 0.11$ & --&$=0^P$ & $<0.06^P$ & $<0.06^P$\\
$e_b$ & $< 0.06$ & $< 0.01$ &-- &$<0.013$ & $< 0.05$ & $< 0.01$ \\
$e_d$ & $< 0.06$ & $<0.024$ &-- &$=0^P$ & $<0.044$ & $< 0.025$ \\
$e_c$ & & $0.36\pm0.12$ & --& $0.27\pm0.04$ &  &$0.28\pm0.02$ \\
\enddata
\tablecomments{B15 - \citet{Becker2015}, K2 TTVs, fixed $P$, $TT$ for each planet, 10 Myr stability enforced.
A16.1 - \citet{Almenara2016}, photodynamical analysis of K2 TTVs and 71 RVs, A16.2 - including stellar models.  S16 - Sinukoff et al. 2016, 118 RVs.  TTVs-Nbody - TTVFast N-body analysis (presented herein), K2 TTVs.  RVs+TTVs - simultaneous analysis of K2 TTVs and 118 RVs.  All upper limits are 95\% confidence. $P$ - results come from a prior.}
\label{tab:rv-ttv}
\end{deluxetable}
Table \ref{tab:rv-ttv} summarizes the relative information in various dynamical analyses of the planet masses and eccentricities.  B15 modeled the K2 TTVs in a 3-planet N-body analysis in which the orbital periods and inital times of transit were fixed, resulting in narrow posteriors for the mass of planet b.  The mass constraint for planet e comes from the choice of prior, rather than the TTVs.  B15 also forward-modeled the system for 10 Myr using Mercury \citep{Chambers1999} to ensure stability, which resulted in the tight eccentricity constraints for the inner planets: $e_k < 0.06$.

\citet[][A16 hereafter]{Almenara2016} did a photodynamical analysis of the K2 TTVs and 71 literature RVs from the PFS and CORALIE spectrographs.  The high signal-to-noise of the transits of planet b allowed them to determine stellar limb-darkening parameters and the planet impact parameter very precisely which, in combination with the small eccentricity of planet b, led to a very precise determination of the stellar density through asterodensity profiling \citep{Kipping2014_asterodensity}.  This led to a model-independent estimate of the planetary masses, presented in column A16.1 of Table \ref{tab:rv-ttv}.  By including constraints from the Dartmouth stellar isochrone models \citep{Dotter2008}, the authors were able to constrain the star and planet masses more precisely, but at the expense of accuracy. The model-dependent star and planet masses are shown in column A16.2.

S17 combined 47 new HIRES RVs with the 71 literature RVs.  While the CORALIE RVs had provided a long baseline enough baseline to detect non-transiting planet c \citep{VanMalle2016} and the PFS RVs had provided sufficiently high precision to detect a marginal RV signal from planet e \citep{Dai2015}, the HIRES RVs provided both a long baseline and high precision in a single dataset.  The HIRES data combined with the other RV datasets resulted in smaller uncertainties for all the planet masses than what had been reported in previous RV studies.

Our TTV-only N-body analysis (TTVs-Nbody) and simultaneous RV and TTV analysis (RVs+TTVs) are shown in Table \ref{tab:rv-ttv}.  The TTVs-Nbody column illustrates how much information is contained in the TTVs.  The RVs+TTVs column illustrates how much information is gained by a joint analysis of the TTVs and RVs.  Our RVs+TTVs analysis confirms the precise values obtained by A16 when they include constraints from stellar models (A16.2).

How much information about planet masses comes from the TTVs?  As discussed in the TTV-only analysis (see Section \ref{sec:TTVs_only}), the TTVs do not provide much information about the transiting planet masses, with the exception of the mass of planet d, which is constrained through the TTVs of planet b.

Simultaneously modeling the TTVs and RVs of planet d yields a more precise determination of the mass of planet d than can be obtained from either analysis alone: the uncertainties shrink from $3.8~\mearth$ (TTVs) and $2.7~\mearth$ (RVs) to $2.0~\mearth$ (TTVs + RVs).  The TTVs provide no information about the mass of planet c, which has a very long period compared to the inner planetary system and thus has no effect on the TTVs.  Thus, the RVs provide the majority of the information about planet masses, although the TTVs contribute substantially to the mass measurement of planet d.

The information about planet eccentricities comes from stability constraints, the TTVs, and the RVs.  The eccentricity of planet e is not constrained by either the TTVs or the RVs, and so its eccentricity varies from 0 to 0.06 (the upper limit from stability requirements).  The RVs constrain $e_b < 0.013$ (95\% confidence, S17).  The TTVs constrain $e_b < 0.02$ (95\% confidence), and the combined RVs+TTVs further constrain $e_b < \eb$ (95\% confidence).  The RVs alone do not provide a strong constraint for the eccentricity of planet d (S17 fixed $e_d=0$).  The TTVs alone constrain $e_d < 0.05$ (95\% confidence), and the combined TTVs+RVs constrain $e_d < 0.025$ (95\% confidence).  Thus, the TTVs provide additional information about the small eccentricities of planets b and d.  The eccentricity of planet c is determined entirely from RVs because the planet is dynamically decoupled from the inner planetary system.

Thus, combining the TTVs and RVs provides more information about masses and eccentricities than either dataset does alone.  We discuss the physical importance of the precise measurements of the masses and eccentricities below.


\subsection{Masses and densities of the planets}
\begin{figure*}
    \centering
    \includegraphics[width=2\columnwidth]{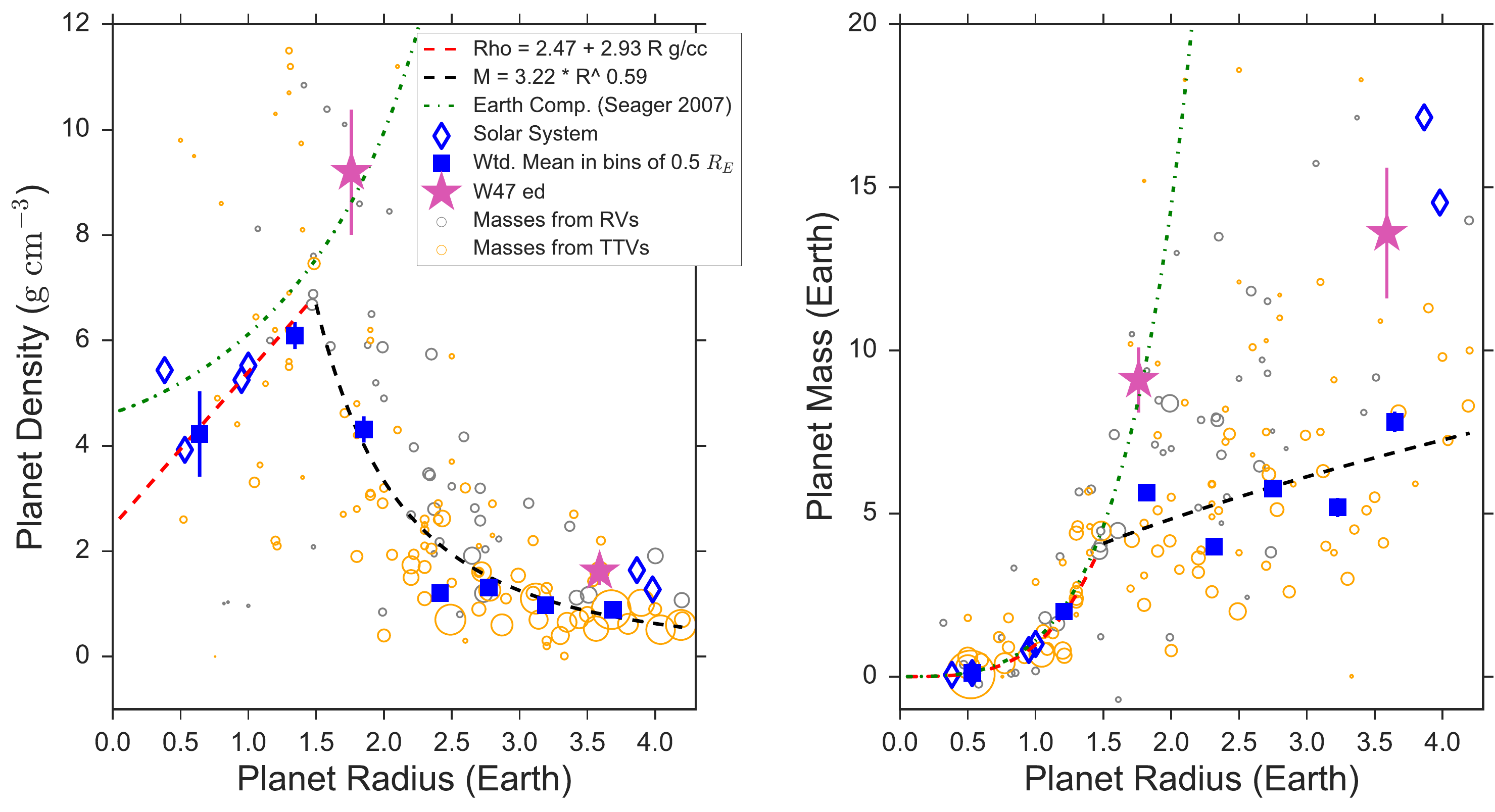}
    \caption{Left: planet density versus planet physical radius for 94 transiting planets smaller than $4.2~\rearth$.  The gray circles have masses determined from RVs; the gold circles have masses determined from TTVs.  The size of the circle corresponds to 1/$\sigma_\rho^2$.  Blue squares show the weighted mean density in bins of $0.5~\rearth$ to guide the eye.  The blue diamonds are the solar system planets.  The red dashed line is an empirical linear fit to planet density versus radius for the exoplanets and solar system planets smaller than $1.5~\rearth$, extended to predict the densities of potentially rocky planets larger than $1.5~\rearth$.  For comparison, we show the predicted density-radius curve for a polytropic equation of state of an Earth-composition planet \citep[][green dotted line]{Seager2007}.  The black line is an empirical power-law fit to planet mass versus radius for planets larger than $1.5~\rearth$.  WASP-47 e ($1.9~\rearth$) sits on the red line and therefore is consistent with a rocky composition, but could also have a thin volatile envelope.  WASP-47 d ($3.7~\rearth$) has a density that requires significant volatiles, including the possibility of high-density volatiles based on the similarity of its bulk properties to Uranus and Neptune.  Right: Same as the left panel, but showing planet mass versus radius, and the circle sizes correspond to $1/\sigma_m^2$.}
    \label{fig:mr_small_W47}
\end{figure*}

WASP-47 is unusual in that the masses of its planets span almost two orders of magnitude.  The low-mass planets e and d are shown in a density-radius and mass-radius plots for small planets (see Figure \ref{fig:mr_small_W47}).  The lowest-mass planet, WASP-47 e, is $\me~\mearth$.  At $1.8~\rearth$, this planet has a density of $\rhoe~\gcc$.  Planets larger than approximately $1.5~\rearth$ are unlikely to be rocky \citep{Weiss2014,Rogers2015}.  Yet, planet e is small and dense enough that a rocky composition is likely based on an extrapolation of the empirical relationship for rocky planets smaller than 1.5 Earth radii \citep{Weiss2014} and theoretical predictions of planet mass and radius for an Earth-like composition \citep{Seager2007}.  However, the density of planet e is also consistent with slightly lower densities that might correspond to a rocky interior overlaid with a thin, low-mass envelope of high mean molecular weight materials.  Such a composition has also been hypothesized for 55 Cancri e, which has a very similar mass, radius, and bulk density to WASP-47 e \citep[][S17]{Lopez2016}.  Like 55 Cnc e, WASP-47 e is also an ultra-short period planet cohabiting an orbital system with giant planets, which underscores the question of how planetary system architecture and planet compositions are related.

By contrast, WASP-47 d, which is $3.6~\rearth$, has a mass of $\md~\mearth$.  This is a slightly higher mass than was reported in S17 because the TTVs add mass information.  The additional information from the TTVs also narrows the mass posterior, shrinking the uncertainty from $2.7$ to $2.0~\mearth$.  The density of WASP-47 d is $\rhod~\gcc$, making it a high-density member of the population of sub-Neptune sized planets with volatile envelopes.  The mass-radius diagram in Figure \ref{fig:mr_small_W47} shows that the mass, radius, and density of WASP-47 d make it one of the most Neptune-like planets discovered to date.  While its composition could be explained by a two-layer model of a H/He envelope atop a silicate-iron core, a Neptune-like composition that includes a thick layer of super-ionic water might also explain the bulk properties of WASP-47 d.

\begin{figure*}
    \centering
    \includegraphics[width=2\columnwidth]{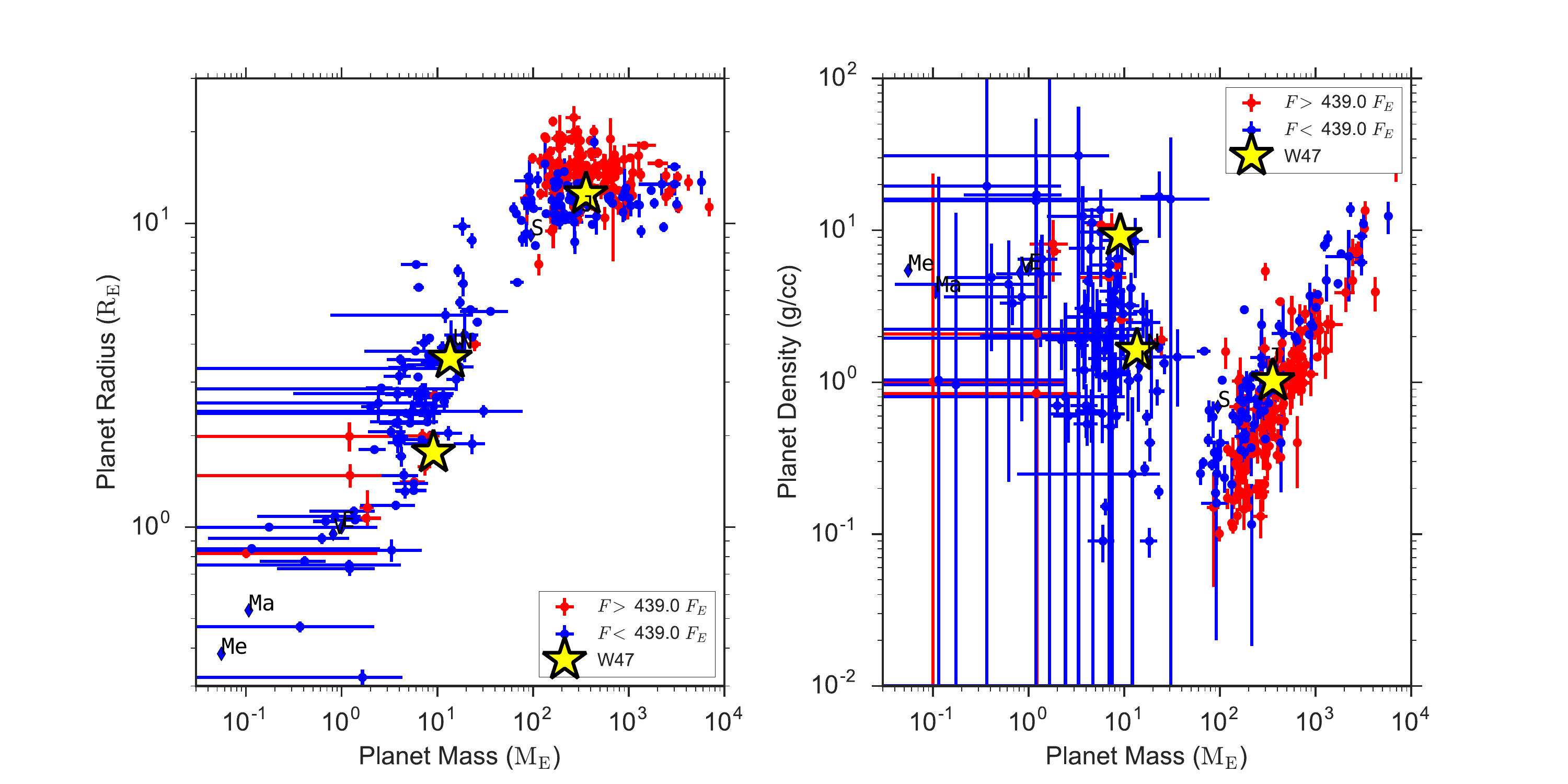}
    \caption{Left: planet radius vs. mass for exoplanets with measured masses and radii, as determined by querying exoplanets.org on 2017/02/19 and including radii and masses from \citep{Hadden2016} and \citep{Gillon2017}.  The sample is divided into those that receive more than the median incident flux ($F > 482 \fearth$, red points) and those that receive less than the median incident flux ($F < 482 \fearth$, blue points). The solar system planets are labeled.  The WASP-47 planets are shown (yellow stars, the mass and radius error bars are smaller than the symbols).  The sub-Neptunes WASP-47 e and d are high-density for their size.  WASP-47 b is a typical-sized hot Jupiter for its mass and incident flux.}
    \label{fig:mrf_w47}
\end{figure*}

WASP-47 b is a Jupiter-mass planet ($\mb~\mearth$) that receives $440\pm70$ times as much incident stellar irradiation as the Earth does.  At $13.11\pm0.89~\rearth$, the planet has a typical density ($\rhob~\gcc$) for its mass and incident stellar flux (see Figure \ref{fig:mrf_w47}), consistent with various theories \citep[][and references therein]{Batygin2011,Fortney2010} that stellar irradiation inflates the planet and/or prevents the planet from cooling.

\subsection{Eccentricities of the planets}
The orbits of the three inner planets are profoundly circular.  B15 found that eccentricities of $<0.06$ were required for stability.  Here, we tighten the eccentricities to $e_b < \eb$ and $e_d < 0.025$ (95\% confidence).  In the highest-eccentricity cases for planets b and d, they tend to be apsidally aligned.  We compute $e_d\mathrm{cos}\omega_d - e_b\mathrm{cos}\omega_b = -0.001\pm0.005$ and $e_d\mathrm{sin}\omega_d - e_b\mathrm{sin}\omega_b = 0.0\pm0.007$.

Although the tidal circularization timescale for planet e is only $\sim10^4-10^5$ years, depending on the tidal Q value for the planet, our N-body analysis revealed that the neighboring giant planet (b) perturbs the eccentricity of planet e on a timescale of 1.26 days.  This timescale happens to be related to the orbital periods of both e and b by $1/P_\mathrm{kick} = (1/P_e + 2/P_b)^{-1}$.  Since there is no commensurability between the orbital periods of b and e, we expect that over long timescales, the kicks from planet b average out, and so the argument of periastron of planet e should not have a preferred value.  However, planet e could still have an average eccentricity that is higher than zero.  Likewise, planet d should not be assumed to have zero eccentricity because it exhibits TTVs.  Because planet d is near the 2:1 resonance with planet b, its argument of periastron circulates at the frequency of the TTV super-period.

\begin{figure}
    \centering
    \includegraphics[width=\columnwidth]{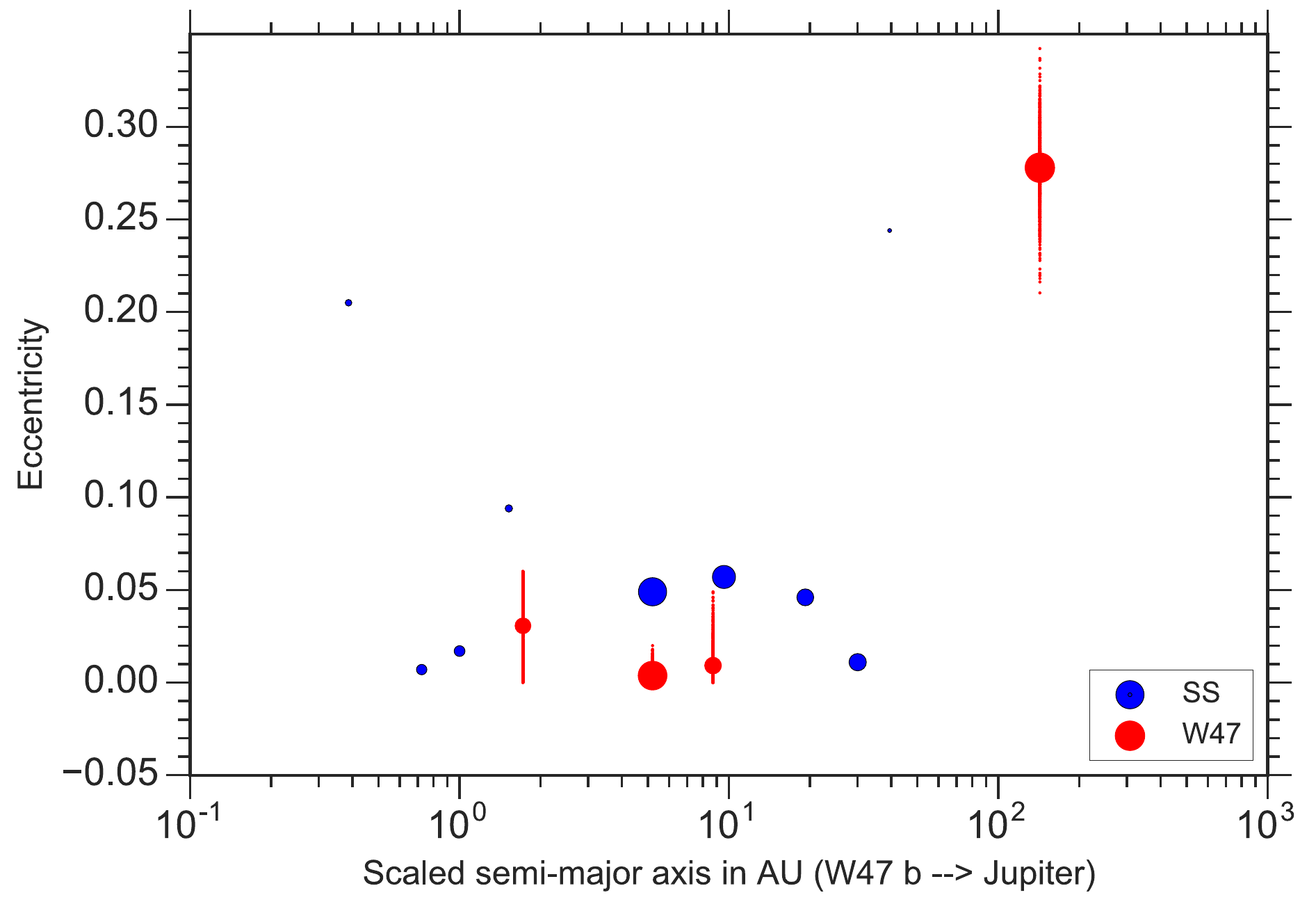}
    \caption{The mean eccentricities of the WASP-47 and solar system planets versus the orbital distances, scaled such that $a_b = a_J$ to facilitate comparison.  The point size corresponds to $\mpl^{1/3}$, to illustrate the mass range while keeping all the planets visible.  The spread of the each WASP-47 planet eccentricity is illustrated with a cloud of 1000 small red points drawn from the eccentricity posterior.  The average eccentricity of the W47 planets is comparable to the average eccentricity of the solar system planets, if we include Pluto.}
    \label{fig:solar_system}
\end{figure}

The very low eccentricities of the WASP-47 planets are remarkably like those of the solar system planets (see Figure \ref{fig:solar_system}).  The average eccentricity of the detected planets in WASP-47 is $ < 0.09$; in the solar system, the average eccentricity of the planets and Pluto is 0.08.

\subsection{Observational Constraints on Formation Theories}
Compact systems like the WASP-47 inner planets might form \textit{in situ} from protoplanetary disks that are more massive than the minimum mass solar nebula \citep{Chiang2013}.  Scenarios in which either gas-poor sub-Neptunes or gas-rich Jupiters form \textit{in situ} have been proposed \citep{Lee2014, Lee2015, Lee2016, Batygin2016}.  If all the WASP-47 planets formed \textit{in situ} from the same nebular material, the challenge is to explain how WASP-47 b managed to achieve runaway growth, whereas its immediate neighbors remained gas-poor.  Whether a Jupiter-mass or sub-Saturn mass planet forms depends on the core mass, atmospheric opacity, and the disk lifetime \citep{Hori&Ikoma2011,Venturini2015}.  Cores of mass $\sim5-15\mearth$ and larger undergo runaway gas accretion \citep[with the exact value of critical core mass depending on the atmospheric opacities, metallicities and the disk gas dissipation timescale;][]{Lee2016,Batygin2016}.  However, gas-poor sub-Neptunes are formed instead of giant planets if the cores form and accrete their envelopes in a gas-poor (transitional) disk.  The formation of sub-Neptune cores by giant impact requires at least four orders of magnitude of gas-depletion with respect to the minimum mass extrasolar nebula \citep{Lee2016}, leaving $\sim0.3~\mearth$ of gas in the disk.  This gas budget is insufficient to form the $\sim300~\mearth$ of gas in WASP-47 b.

A modification of in situ formation is inside-out growth via pebble accretion, in which pebbles are transported inward through the disk until they are stopped by a pressure maximum at the boundary between the magneto-rotational instability (MRI) zone and the magnetic dead zone \citep{Chatterjee2014}.  At this boundary, the pebbles may become Toomre unstable or may coalesce via core accretion.  As the gas disk clears and the boundary between the MRI and dead zone moves outward, the site of planet formation gradually moves out through the disk.  Because the hot Jupiter is situated between two sub-Neptunes, the accretion rate of the disk would likely need to increase by roughly an order of magnitude, and then decrease again, to explain the high mass of the Jupiter compared to its neighbors.  Furthermore, the efficacy of forming hot Jupiters via inside-out planet formation has not been well studied.

Alternatively, the planets could have formed elsewhere in the disk and then migrated via interactions with the disk to their present locations. In both Type I and Type II migration, the gas disk damps planet eccentricity, allowing planets to maintain circular orbits in a manner consistent with the nearly circular orbits of the three inner planets.  However, slow migration can trap the planets in mean motion resonances.  While WASP-47 b and d are near the 2:1 mean motion resonance, they are not trapped.  Thus, their migration history must either include a mechanism to prevent planets b and d from entering the 2:1 resonance, or remove them from the resonance \citep[e.g.,][]{Adams2008,Goldreich&Schlichting2014}.  In particular, \citet{Deck&Batygin2015} find that if the inner planet of a pair near a first-order mean motion resonance is the more massive (as is the case for WASP-47 b and d), escape from resonance is unlikely.  Furthermore, migration in the disk does not explain the eccentricity of planet c (\ec).  Also, planets e and d would need to migrate through the disk without accreting gas.

Planet-planet scattering and Kozai-Lidov oscillations are big-body (as opposed to gas-and-dust) migration mechanisms. After the gas disk has dissipated, these mechanisms can introduce moderate to high eccentricities in the orbits, potentially accounting for the moderate eccentricity of planet c.  Kozai-Lidov oscillations are initiated only when two planets have mutual inclinations of at least $40^\circ$.  The planets swap angular momentum, causing dramatic variations in the inclinations and eccentricities of the planets over time \citep{Kozai1962, Lidov1962}.  The strongest lines of evidence for past  Kozai-Lidov interactions would be (1) an observed large mutual inclination between planet c and the inner solar system, and/or (2) a non-zero obliquity between planet b and the stellar spin axis. \citet{Sanchis-Ojeda2015} ruled out a significant spin-orbit misalignment for planet b, but the degeneracy between stellar $\vsini$ and the planet-star obliquity allows moderate misalignments ($|\lambda| < 48^\circ$ with 1$\sigma$ confidence).  If planet c is coplanar and the spin-orbit alignments of the planets are small, a coplanar, high-eccentricity migration (HEM) scenario such as the one proposed in \citet{Petrovich2015} might best explain the present locations of the planets.  However, the current nearly-circular, coplanar orbits of the inner planetary system are not consistent with a past modest eccentricity and/or inclination for planet b.  Recall from the stability analysis in B15 that the nearly all of the scenarios in which the eccentricity of planet b exceeds 0.06 become unstable within 10 Myr.  Therefore, although high eccentricity migration mechanisms can explain the present positions of planets b and c, such mechanisms would have destroyed the small, close-in planets.  Indeed, simulations of giant planets migrating inward find that the giant planets collide with low-mass inner planets in both high-eccentricity \citep{Mustill2015} and low-eccentricity \citep{Batygin2015_GrandAttack} scenarios.

\subsection{The Two Stage Planet Formation Hypothesis}
One way to explain the present orbits and compositions of the WASP-47 planets is that the planets did not all form at the same time.  As a point of reference, consider the solar system.  In our own solar system, the giant planets must have formed early (within $~$1 to 10 Myr), when the proto-planetary disk was still gas-rich \citep{dePater&Lissauer2001}.  In contrast, the formation of terrestrial-mass planets by planetesimal accretion can take as long as $\sim10^8$ years in N-body simulations and is more efficient in gas-poor disks where the planetesimal eccentricities can grow, leading to more collisions \citep{Lissauer1987, Pollack1996, Ida&Lin2004,Lee2016}.  Thus, it is possible that the solar system giant planets formed and migrated to their present locations before the terrestrial planets formed.  The Nice Model demonstrates that early formation and migration of Jupiter and Saturn can reproduce the current orbital architectures of the gas giants and ice giants while also accounting for the period of Late Heavy Bombardment on the terrestrial planets, and Trojan asteroids \citep{Gomes2005,Tsiganis2005, Morbidelli2005}.  The Grand Tack Model shows that a reversal in the direction of Jupiter's migration due to torques from Saturn can explain the small size of Mars and detailed compositional features of the asteroids \citep{Walsh2011, Morbidelli2012}.  Both of these models feature two stages of planet formation: 
\begin{enumerate}
\item Early giant planet growth combined with disk and/or planet-induced migration produces the current compositions of giant planets and sculpts the radial distribution of planetesimals.  
\item When or after the gas disk clears, planetesimal accretion proceeds in the sculpted planetesimal disk, resulting in the formation of low-mass, predominantly rocky planets.
\end{enumerate}

Likewise, the formation and evolution of the planets in WASP-47 might be best explained by a two-stage process.  The giant planets in the WASP-47 system might have formed early in the gas-rich disk and exchanged energy with each other, additional bodies, or the disk to migrate to their present locations.  If the gas disk cleared while WASP-47 b and c were finishing their migration, the final years of the giant planets' migration could have influenced the solid material in the disk, exciting protoplanetary solids to higher eccentricities and inducing a second stage of core accretion.  Such an epoch could have formed WASP-47 e and d, and perhaps additional yet-undetected low-mass planets dominated by rocky interiors.

\subsection{Predictions of Two-Stage Planet Formation}
In WASP-47, the orbital architecture and compositions of the low-mass planets provide insight into the formation history of the giant planets.  Because the low-mass planets need a small amount of gas to form (about $0.3~\mearth$), their formation must occur before the gas disk completely dissipates.  This puts a deadline on when planet b can finish its migration; it must park at its current location before planets e and d form (otherwise it destroys them).  In other words, planet b must arrive at its present location before the gas disk dissipates.

Because a gas disk damps planet eccentricities on the timescale of $10^3$ years \citep{Papaloizou&Larwood2000}, high-eccentricity migration cannot proceed in a gas disk.  Furthermore, the timescale for type I migration in a gas disk is much faster than the timescale for HEM, and so gas disk migration dominates while the gas disk is present.  If we consider a mostly gas-depleted disk, we can try to migrate WASP-47 b inward via HEM in a short window of time before the disk completely dissipates.  However, some force must circularize planet b's orbit quickly enough to allow the neighboring small planets to form.  In the absence of gas, the only circularizing force is tides raised on the planet by the star.  The timescale for eccentricity damping due to tides \citep[][Equation 4.198]{Murray&DermottCh8} is:
\begin{equation}
\tau_e = -\frac{e}{\dot{e}} = \frac{4}{63}\frac{\mpl}{\mstar}\big(\frac{a}{\rpl}\big)^5 \frac{\mu_\mathrm{pl} Q_\mathrm{pl}}{2\pi/P}.
\end{equation}
where $\mu_\mathrm{pl}$ is the ratio of the elastic to gravitational forces in the planet $\approx(10^4~\mathrm{km}/\rpl)^2$, and $Q_\mathrm{pl}$ is the tidal dissipation factor of the planet.  For $a=0.05$ AU and $Q_\mathrm{pl} = 10^{6.5}$ \citep[typical for hot Jupiters,][]{Jackson2008}, the circularization time is $\tau_e \approx 10^8~\mathrm{yr}$, which is an order of magnitude longer than the disk lifetime.  There is not enough time to move WASP-47 b to its present position via HEM and then circularize its orbit before the little planets form.  Therefore, a history of HEM is highly unlikely for planet b.

On the other hand, planet b could also have formed \textit{in situ} or nearly \textit{in situ} in the gas-rich disk.  \citet{Antonini2016} find through N-body simulations that, for the majority of observed Jupiter pairs (i.e., systems with one Jupiter-mass planet inside 1 A.U. and one Jupiter-mass planet outside 1 A.U.), the inner planet is unlikely to have formed beyond 1 A.U.  Thus, it seems probable that WASP-47 b formed inside 1 A.U., and thus inside the snow line.

Since planet c has modest eccentricity (\ec), it almost certainly had some sort of planet-planet interaction in the past.  Although it is possible for instabilities in the gas disk to excite giant planet eccentricities, this effect only works for $e < 0.1$ and therefore cannot explain the eccentricity of planet c \citep{Duffell&Chiang2015}.  As described above, planet b is not a likely candidate for pumping the eccentricity of planet c.  Thus, additional massive bodies that exchanged angular momentum with planet c in the past might still be present in the WASP-47 system.  Alternatively, planet c might have ejected whatever companion pumped its eccentricity.

Based on the above arguments, evidence of two-stage planet formation in WASP-47 could include:
\begin{enumerate}
\item a C/O ratio for WASP-47 b, e and d consistent with formation inside the snowline,
\item alignment of the stellar spin axis with the orbits of the transiting planets, consistent with a coplanar, mostly circular orbital history for planet b,
\item an additional massive planet/sub-stellar/stellar companion with an eccentric orbit that exchanged angular momentum with planet c in the past.
\end{enumerate}

A two-stage formation mechanism might also best explain the current architectures of other multi-planet systems with diverse planet compositions, such as 55 Cancri and Kepler-89 (KOI-94).  An improved census of giant planets accompanying the low-mass planetetary systems discovered by \textit{Kepler} will enable studies of the occurrence of planetary systems with diverse compositions like our solar system, illuminating the dominant physical processes in their formation.

Continued observations of this system and other systems with diverse planet compositions from the James Webb Space Telescope, the WFIRST mission, and other facilities will provide evidence for or against a two-stage planet formation scenario.




\section{Conclusion}
\label{sec:conclusion}
We combine 118 RVs and 108 K2 TTVs to present some of the most precise masses, densities, and orbital dynamics of the WASP-47 planetary system to date.  For the transiting inner planetary system, we obtain $M_e = \me~\mearth$ ($\rho_e = \rhoe~\gcc$), $M_b = \mb~\mearth$ ($\rho_b = \rhob~\gcc$), and $M_d = \md~\mearth$ ($\rho_d = \rhod~\gcc$).

We place tight upper limits on the eccentricities of the three inner planets: $e_e < 0.06$, $e_b < \eb$, and $e_d < \ed$ (95\% confidence).  The mean eccentricity of the WASP-47 planets (0.09) is very similar to the mean eccentricity of the solar system planets plus Pluto (0.08).

Of the four planets, only WASP-47 e has bulk properties consistent with a rocky composition.  WASP-47 d is a sub-Neptune sized planet that is just a little smaller than Neptune but has the density of Neptune, meaning that its bulk properties consistent with a Neptune-like composition.  However, WASP-47 d could also be a rocky interior overlaid with a hydrogen-rich, water-poor envelope.  WASP-47 b is a hot Jupiter with a typical size for its mass and incident stellar flux.

The non-resonant architecture, diverse masses, and profoundly circular orbits in the WASP-47 resemble our own solar system.  We briefly review the highlights of \textit{in situ} planet formation, inside-out planet formation, disk migration, and high-eccentricity migration, noting that none of these mechanisms alone can reproduce all the physical attributes of the WASP-47 system.  We propose that WASP-47, like the solar system, formed in two stages.  In stage one, the giant planets formed in a gas-rich disk and migrated via disk migration to their present locations.  In stage two, the high-density sub-Neptunes formed \textit{in situ} in a gas-poor environment.

\acknowledgments
L.~M.~W.\ acknowledges the Trottier Family Foundation for their support. K.~D.\ acknowledges the support of the JCPA fellowship at Caltech.  E.~S.\ is supported by a post-graduate scholarship from the Natural Sciences and Engineering Research Council of Canada.  E.~A.\ acknowledges support from NASA grants
NNX13AF20G, NNX13AF62G, and NASA Astrobiology Institutes Virtual Planetary
Laboratory, supported by NASA under cooperative agreement NNH05ZDA001C.  A.~W.~H.\ acknowledges support from a NASA Astrophysics Data Analysis Program grant, support from the K2 Guest Observer Program, and a NASA Key Strategic Mission Support Project.  We thank Simon Walker for his help and generosity in using ttvfast-python.  The authors thank Geoff Marcy, Eva Culakova, Daniel Fabrycky, Jack Lissauer, Jason Rowe, the Kepler-TTV working group, and the KITP Planet Formation and Dynamics workshop for useful discussions.  We thank NASA and the Kepler team for the outstanding photometry that contributed to this paper.  The authors wish to extend special thanks to those of Hawaiian ancestry on whose sacred mountain of Maunakea we are privileged to be guests. Without their generous hospitality, the Keck observations analyzed herein would not have been possible.
\facilities{Keck:I (HIRES), Kepler}

\software{TTVFaster (Agol \& Deck 2016a,b), TTVFast (Deck et al. 2014), ttvfast-python (https://github.com/mindriot101/ttvfast-python), emcee (Foreman-Mackey et al. 2013), lmfit (Newville et al. 2014), radvel (https://github.com/California-Planet-Search/radvel)}

\bibliographystyle{apj}
\bibliography{references}

\begin{thebibliography}{}
\expandafter\ifx\csname natexlab\endcsname\relax\def\natexlab#1{#1}\fi

\bibitem[{{Adams} {et~al.}(2008){Adams}, {Laughlin}, \& {Bloch}}]{Adams2008}
{Adams}, F.~C., {Laughlin}, G., \& {Bloch}, A.~M. 2008, \apj, 683, 1117

\bibitem[{{Agol} \& {Deck}(2016{\natexlab{a}})}]{Agol2016}
{Agol}, E., \& {Deck}, K. 2016{\natexlab{a}}, \apj, 818, 177

\bibitem[{{Agol} \& {Deck}(2016{\natexlab{b}})}]{Agol2016TTVFaster}
---. 2016{\natexlab{b}}, {TTVFaster: First order eccentricity transit timing
  variations (TTVs)}, Astrophysics Source Code Library, ascl:1604.012

\bibitem[{{Almenara} {et~al.}(2016){Almenara}, {D{\'{\i}}az}, {Bonfils}, \&
  {Udry}}]{Almenara2016}
{Almenara}, J.~M., {D{\'{\i}}az}, R.~F., {Bonfils}, X., \& {Udry}, S. 2016,
  \aap, 595, L5

\bibitem[{{Antonini} {et~al.}(2016){Antonini}, {Hamers}, \&
  {Lithwick}}]{Antonini2016}
{Antonini}, F., {Hamers}, A.~S., \& {Lithwick}, Y. 2016, \aj, 152, 174

\bibitem[{Batalha {et~al.}(2013)Batalha, Rowe, Bryson, Barclay, Burke,
  Caldwell, Christiansen, Mullally, Thompson, Brown, Dupree, Fabrycky, Ford,
  Fortney, Gilliland, Isaacson, Latham, Marcy, Quinn, Ragozzine, Shporer,
  Borucki, Ciardi, Gautier, Haas, Jenkins, Koch, Lissauer, Rapin, Basri, Boss,
  Buchhave, Carter, Charbonneau, Christensen-Dalsgaard, Clarke, Cochran,
  Demory, Desert, Devore, Doyle, Esquerdo, Everett, Fressin, Geary, Girouard,
  Gould, Hall, Holman, Howard, Howell, Ibrahim, Kinemuchi, Kjeldsen, Klaus, Li,
  Lucas, Meibom, Morris, Pr\v{s}a, Quintana, Sanderfer, Sasselov, Seader,
  Smith, Steffen, Still, Stumpe, Tarter, Tenenbaum, Torres, Twicken, Uddin,
  {Van Cleve}, Walkowicz, \& Welsh}]{Batalha2013}
Batalha, N.~M., Rowe, J.~F., Bryson, S.~T., {et~al.} 2013, The Astrophysical
  Journal Supplement Series, 204, 24

\bibitem[{{Batygin} {et~al.}(2016){Batygin}, {Bodenheimer}, \&
  {Laughlin}}]{Batygin2016}
{Batygin}, K., {Bodenheimer}, P.~H., \& {Laughlin}, G.~P. 2016, \apj, 829, 114

\bibitem[{Batygin \& Laughlin(2015)}]{Batygin2015_GrandAttack}
Batygin, K., \& Laughlin, G. 2015, Proceedings of the National Academy of
  Sciences, 112, 4214

\bibitem[{Batygin {et~al.}(2011)Batygin, Stevenson, \&
  Bodenheimer}]{Batygin2011}
Batygin, K., Stevenson, D.~J., \& Bodenheimer, P.~H. 2011, \apj, 738, 1

\bibitem[{{Becker} {et~al.}(2015){Becker}, {Vanderburg}, {Adams}, {Rappaport},
  \& {Schwengeler}}]{Becker2015}
{Becker}, J.~C., {Vanderburg}, A., {Adams}, F.~C., {Rappaport}, S.~A., \&
  {Schwengeler}, H.~M. 2015, \apjl, 812, L18

\bibitem[{Borucki {et~al.}(2010)Borucki, Koch, Brown, Basri, Batalha, Caldwell,
  Cochran, Dunham, {Gautier III}, Geary, Gilliland, Howell, Jenkins, Latham,
  Lissauer, Marcy, Monet, Rowe, \& Sasselov}]{Borucki2010}
Borucki, W.~J., Koch, D.~G., Brown, T.~M., {et~al.} 2010, \apjl, 713, L126

\bibitem[{Brooks \& Gelman(1998)}]{Brooks1998}
Brooks, S.~P., \& Gelman, A. 1998, Journal of computational and graphical
  statistics, 7, 434

\bibitem[{{Bryan} {et~al.}(2016){Bryan}, {Knutson}, {Howard}, {Ngo}, {Batygin},
  {Crepp}, {Fulton}, {Hinkley}, {Isaacson}, {Johnson}, {Marcy}, \&
  {Wright}}]{Bryan2016}
{Bryan}, M.~L., {Knutson}, H.~A., {Howard}, A.~W., {et~al.} 2016, \apj, 821, 89

\bibitem[{{Burke} {et~al.}(2015){Burke}, {Christiansen}, {Mullally}, {Seader},
  {Huber}, {Rowe}, {Coughlin}, {Thompson}, {Catanzarite}, {Clarke}, {Morton},
  {Caldwell}, {Bryson}, {Haas}, {Batalha}, {Jenkins}, {Tenenbaum}, {Twicken},
  {Li}, {Quintana}, {Barclay}, {Henze}, {Borucki}, {Howell}, \&
  {Still}}]{Burke2015}
{Burke}, C.~J., {Christiansen}, J.~L., {Mullally}, F., {et~al.} 2015, \apj,
  809, 8

\bibitem[{Chambers(1999)}]{Chambers1999}
Chambers, J.~E. 1999, \mnras, 304, 793

\bibitem[{{Chatterjee} \& {Tan}(2014)}]{Chatterjee2014}
{Chatterjee}, S., \& {Tan}, J.~C. 2014, \apj, 780, 53

\bibitem[{Chiang \& Laughlin(2013)}]{Chiang2013}
Chiang, E., \& Laughlin, G. 2013, Monthly Notices of the Royal Astronomical
  Society, 431, 3444

\bibitem[{{Dai} {et~al.}(2015){Dai}, {Winn}, {Arriagada}, {Butler}, {Crane},
  {Johnson}, {Shectman}, {Teske}, {Thompson}, {Vanderburg}, \&
  {Wittenmyer}}]{Dai2015}
{Dai}, F., {Winn}, J.~N., {Arriagada}, P., {et~al.} 2015, \apjl, 813, L9

\bibitem[{{de Pater} \& {Lissauer}(2001)}]{dePater&Lissauer2001}
{de Pater}, I., \& {Lissauer}, J.~J. 2001, {Planetary Sciences}, 544

\bibitem[{Deck {et~al.}(2014)Deck, Agol, Holman, \& Nesvorn\'{y}}]{Deck2014}
Deck, K.~M., Agol, E., Holman, M.~J., \& Nesvorn\'{y}, D. 2014, The
  Astrophysical Journal, 787, 132

\bibitem[{{Deck} \& {Batygin}(2015)}]{Deck&Batygin2015}
{Deck}, K.~M., \& {Batygin}, K. 2015, \apj, 810, 119

\bibitem[{{Dotter} {et~al.}(2008){Dotter}, {Chaboyer}, {Jevremovi{\'c}},
  {Kostov}, {Baron}, \& {Ferguson}}]{Dotter2008}
{Dotter}, A., {Chaboyer}, B., {Jevremovi{\'c}}, D., {et~al.} 2008, \apjs, 178,
  89

\bibitem[{{Dressing} \& {Charbonneau}(2015)}]{Dressing2015Mdwarfs}
{Dressing}, C.~D., \& {Charbonneau}, D. 2015, \apj, 807, 45

\bibitem[{{Duffell} \& {Chiang}(2015)}]{Duffell&Chiang2015}
{Duffell}, P.~C., \& {Chiang}, E. 2015, \apj, 812, 94

\bibitem[{{Fabrycky} {et~al.}(2014){Fabrycky}, {Lissauer}, {Ragozzine}, {Rowe},
  {Steffen}, {Agol}, {Barclay}, {Batalha}, {Borucki}, {Ciardi}, {Ford},
  {Gautier}, {Geary}, {Holman}, {Jenkins}, {Li}, {Morehead}, {Morris},
  {Shporer}, {Smith}, {Still}, \& {Van Cleve}}]{Fabrycky2014}
{Fabrycky}, D.~C., {Lissauer}, J.~J., {Ragozzine}, D., {et~al.} 2014, \apj,
  790, 146

\bibitem[{Fang \& Margot(2012)}]{Fang2012}
Fang, J., \& Margot, J.-L. 2012, \apj, 761, 92

\bibitem[{{Foreman-Mackey} {et~al.}(2013){Foreman-Mackey}, {Hogg}, {Lang}, \&
  {Goodman}}]{Foreman-Mackey2013}
{Foreman-Mackey}, D., {Hogg}, D.~W., {Lang}, D., \& {Goodman}, J. 2013, \pasp,
  125, 306

\bibitem[{{Fortney} \& {Nettelmann}(2010)}]{Fortney2010}
{Fortney}, J.~J., \& {Nettelmann}, N. 2010, \ssr, 152, 423

\bibitem[{Fressin {et~al.}(2013)Fressin, Torres, Charbonneau, Bryson,
  Christiansen, Dressing, Jenkins, Walkowicz, \& Batalha}]{Fressin2013}
Fressin, F., Torres, G., Charbonneau, D., {et~al.} 2013, The Astrophysical
  Journal, 766, 81

\bibitem[{Gelman \& Rubin(1992)}]{Gelman1992}
Gelman, A., \& Rubin, D.~B. 1992, Statist. Sci., 7, 457

\bibitem[{Gillon {et~al.}(2017)Gillon, Triaud, Demory, Jehin, Agol, Deck,
  Lederer, Wit, Burdanov, Ingalls, Bolmont, Leconte, Raymond, Selsis, Turbet,
  Barkaoui, Burgasser, Burleigh, Carey, Chaushev, Copperwheat, Delrez,
  Fernandes, Holdsworth, Kotze, Grootel, Almleaky, Benkhaldoun, Magain, \&
  Queloz}]{Gillon2017}
Gillon, M., Triaud, A. H. M.~J., Demory, B.-O., {et~al.} 2017, Nature, 542, 456

\bibitem[{Gladman(1993)}]{Gladman1993}
Gladman, B. 1993, Icarus, 106, 247

\bibitem[{{Goldreich} \& {Schlichting}(2014)}]{Goldreich&Schlichting2014}
{Goldreich}, P., \& {Schlichting}, H.~E. 2014, \aj, 147, 32

\bibitem[{{Gomes} {et~al.}(2005){Gomes}, {Levison}, {Tsiganis}, \&
  {Morbidelli}}]{Gomes2005}
{Gomes}, R., {Levison}, H.~F., {Tsiganis}, K., \& {Morbidelli}, A. 2005, \nat,
  435, 466

\bibitem[{{Hadden} \& {Lithwick}(2016)}]{Hadden2016}
{Hadden}, S., \& {Lithwick}, Y. 2016, ArXiv e-prints, arXiv:1611.03516

\bibitem[{{Hellier} {et~al.}(2012){Hellier}, {Anderson}, {Collier Cameron},
  {Doyle}, {Fumel}, {Gillon}, {Jehin}, {Lendl}, {Maxted}, {Pepe}, {Pollacco},
  {Queloz}, {S{\'e}gransan}, {Smalley}, {Smith}, {Southworth}, {Triaud},
  {Udry}, \& {West}}]{Hellier2012}
{Hellier}, C., {Anderson}, D.~R., {Collier Cameron}, A., {et~al.} 2012, \mnras,
  426, 739

\bibitem[{{Hori} \& {Ikoma}(2011)}]{Hori&Ikoma2011}
{Hori}, Y., \& {Ikoma}, M. 2011, \mnras, 416, 1419

\bibitem[{Howard {et~al.}(2012)Howard, Marcy, Bryson, Jenkins, Rowe, Batalha,
  Borucki, Koch, Dunham, {Gautier III}, {Van Cleve}, Cochran, Latham, Lissauer,
  Torres, Brown, Gilliland, Buchhave, Caldwell, Christensen-Dalsgaard, Ciardi,
  Fressin, Haas, Howell, Kjeldsen, Seager, Rogers, Sasselov, Steffen, Basri,
  Charbonneau, Christiansen, Clarke, Dupree, Fabrycky, Fischer, Ford, Fortney,
  Tarter, Girouard, Holman, Johnson, Klaus, Machalek, Moorhead, Morehead,
  Ragozzine, Tenenbaum, Twicken, Quinn, Isaacson, Shporer, Lucas, Walkowicz,
  Welsh, Boss, Devore, Gould, Smith, Morris, Prsa, Morton, Still, Thompson,
  Mullally, Endl, \& MacQueen}]{Howard2012}
Howard, A.~W., Marcy, G.~W., Bryson, S.~T., {et~al.} 2012, \apjs, 201, 15

\bibitem[{{Ida} \& {Lin}(2004)}]{Ida&Lin2004}
{Ida}, S., \& {Lin}, D.~N.~C. 2004, \apj, 604, 388

\bibitem[{{Jackson} {et~al.}(2008){Jackson}, {Greenberg}, \&
  {Barnes}}]{Jackson2008}
{Jackson}, B., {Greenberg}, R., \& {Barnes}, R. 2008, \apj, 678, 1396

\bibitem[{{Kipping}(2014)}]{Kipping2014_asterodensity}
{Kipping}, D.~M. 2014, \mnras, 440, 2164

\bibitem[{{Koch} {et~al.}(2010){Koch}, {Borucki}, {Rowe}, {Batalha}, {Brown},
  {Caldwell}, {Caldwell}, {Cochran}, {DeVore}, {Dunham}, {Dupree}, {Gautier},
  {Geary}, {Gilliland}, {Howell}, {Jenkins}, {Latham}, {Lissauer}, {Marcy},
  {Morrison}, \& {Tarter}}]{Koch2010}
{Koch}, D.~G., {Borucki}, W.~J., {Rowe}, J.~F., {et~al.} 2010, \apjl, 713, L131

\bibitem[{{Kozai}(1962)}]{Kozai1962}
{Kozai}, Y. 1962, \aj, 67, 591

\bibitem[{{Lee} \& {Chiang}(2015)}]{Lee2015}
{Lee}, E.~J., \& {Chiang}, E. 2015, \apj, 811, 41

\bibitem[{Lee \& Chiang(2016)}]{Lee2016}
Lee, E.~J., \& Chiang, E. 2016, The Astrophysical Journal, 817, 90

\bibitem[{{Lee} {et~al.}(2014){Lee}, {Chiang}, \& {Ormel}}]{Lee2014}
{Lee}, E.~J., {Chiang}, E., \& {Ormel}, C.~W. 2014, \apj, 797, 95

\bibitem[{{Lidov}(1962)}]{Lidov1962}
{Lidov}, M.~L. 1962, \planss, 9, 719

\bibitem[{{Lissauer}(1987)}]{Lissauer1987}
{Lissauer}, J.~J. 1987, \icarus, 69, 249

\bibitem[{Lissauer {et~al.}(2012)Lissauer, Marcy, Rowe, Bryson, Adams,
  Buchhave, Ciardi, Cochran, Fabrycky, Ford, Fressin, Geary, Gilliland, Holman,
  Howell, Jenkins, Kinemuchi, Koch, Morehead, Ragozzine, Seader, Tanenbaum,
  Torres, \& Twicken}]{Lissauer2012}
Lissauer, J.~J., Marcy, G.~W., Rowe, J.~F., {et~al.} 2012, \apj, 750, 112

\bibitem[{Lissauer {et~al.}(2014)Lissauer, Marcy, Bryson, Rowe, Jontof-Hutter,
  Agol, Borucki, Carter, Ford, Gilliland, Kolbl, Star, Steffen, \&
  Torres}]{Lissauer2014}
Lissauer, J.~J., Marcy, G.~W., Bryson, S.~T., {et~al.} 2014, \apj, 784, 44

\bibitem[{{Lithwick} {et~al.}(2012){Lithwick}, {Xie}, \& {Wu}}]{Lithwick2012}
{Lithwick}, Y., {Xie}, J., \& {Wu}, Y. 2012, \apj, 761, 122

\bibitem[{{Lopez}(2016)}]{Lopez2016}
{Lopez}, E.~D. 2016, ArXiv e-prints, arXiv:1610.01170

\bibitem[{{Lovis} \& {Fischer}(2010)}]{Seager_exoplanets2010_lovis}
{Lovis}, C., \& {Fischer}, D. 2010, {Radial Velocity Techniques for
  Exoplanets}, ed. S.~{Seager}, 27--53

\bibitem[{{Morbidelli} {et~al.}(2005){Morbidelli}, {Levison}, {Tsiganis}, \&
  {Gomes}}]{Morbidelli2005}
{Morbidelli}, A., {Levison}, H.~F., {Tsiganis}, K., \& {Gomes}, R. 2005, \nat,
  435, 462

\bibitem[{{Morbidelli} {et~al.}(2012){Morbidelli}, {Lunine}, {O'Brien},
  {Raymond}, \& {Walsh}}]{Morbidelli2012}
{Morbidelli}, A., {Lunine}, J.~I., {O'Brien}, D.~P., {Raymond}, S.~N., \&
  {Walsh}, K.~J. 2012, Annual Review of Earth and Planetary Sciences, 40, 251

\bibitem[{{Murray} \& {Correia}(2010)}]{Seager_exoplanets2010_murray}
{Murray}, C.~D., \& {Correia}, A.~C.~M. 2010, {Keplerian Orbits and Dynamics of
  Exoplanets}, ed. S.~{Seager}, 15--23

\bibitem[{Murray \& Dermott(2000)}]{Murray&DermottCh8}
Murray, C.~D., \& Dermott, S.~F. 2000, in Solar System Dynamics (Cambridge
  University Press), 321--408, cambridge Books Online

\bibitem[{{Mustill} {et~al.}(2015){Mustill}, {Davies}, \&
  {Johansen}}]{Mustill2015}
{Mustill}, A.~J., {Davies}, M.~B., \& {Johansen}, A. 2015, \apj, 808, 14

\bibitem[{{Neveu-VanMalle} {et~al.}(2016){Neveu-VanMalle}, {Queloz},
  {Anderson}, {Brown}, {Collier Cameron}, {Delrez}, {D{\'{\i}}az}, {Gillon},
  {Hellier}, {Jehin}, {Lister}, {Pepe}, {Rojo}, {S{\'e}gransan}, {Triaud},
  {Turner}, \& {Udry}}]{VanMalle2016}
{Neveu-VanMalle}, M., {Queloz}, D., {Anderson}, D.~R., {et~al.} 2016, \aap,
  586, A93

\bibitem[{Newville {et~al.}(2014)Newville, Stensitzki, Allen, \&
  Ingargiola}]{Newville2014}
Newville, M., Stensitzki, T., Allen, D.~B., \& Ingargiola, A. 2014, {LMFIT:
  Non-Linear Least-Square Minimization and Curve-Fitting for Python¶},
  doi:10.5281/zenodo.11813

\bibitem[{{Papaloizou} \& {Larwood}(2000)}]{Papaloizou&Larwood2000}
{Papaloizou}, J.~C.~B., \& {Larwood}, J.~D. 2000, \mnras, 315, 823

\bibitem[{Petigura {et~al.}(2013{\natexlab{a}})Petigura, Howard, \&
  Marcy}]{Petigura2013}
Petigura, E.~A., Howard, A.~W., \& Marcy, G.~W. 2013{\natexlab{a}}, Proceedings
  of the National Academy of Sciences, 110, 19273

\bibitem[{Petigura {et~al.}(2013{\natexlab{b}})Petigura, Marcy, \&
  Howard}]{Petigura2013_jan}
Petigura, E.~A., Marcy, G.~W., \& Howard, A.~W. 2013{\natexlab{b}}, \apj, 770,
  69

\bibitem[{{Petrovich}(2015)}]{Petrovich2015}
{Petrovich}, C. 2015, \apj, 805, 75

\bibitem[{{Pollack} {et~al.}(1996){Pollack}, {Hubickyj}, {Bodenheimer},
  {Lissauer}, {Podolak}, \& {Greenzweig}}]{Pollack1996}
{Pollack}, J.~B., {Hubickyj}, O., {Bodenheimer}, P., {et~al.} 1996, \icarus,
  124, 62

\bibitem[{Rogers(2015)}]{Rogers2015}
Rogers, L.~A. 2015, \apj, 801, 41

\bibitem[{{Rowe} {et~al.}(2014){Rowe}, {Bryson}, {Marcy}, {Lissauer},
  {Jontof-Hutter}, {Mullally}, {Gilliland}, {Issacson}, {Ford}, {Howell},
  {Borucki}, {Haas}, {Huber}, {Steffen}, {Thompson}, {Quintana}, {Barclay},
  {Still}, {Fortney}, {Gautier}, {Hunter}, {Caldwell}, {Ciardi}, {Devore},
  {Cochran}, {Jenkins}, {Agol}, {Carter}, \& {Geary}}]{Rowe2014}
{Rowe}, J.~F., {Bryson}, S.~T., {Marcy}, G.~W., {et~al.} 2014, \apj, 784, 45

\bibitem[{Sanchis-Ojeda {et~al.}(2013)Sanchis-Ojeda, Rappaport, Winn, Levine,
  Kotson, Latham, \& Buchhave}]{Sanchis-Ojeda2013}
Sanchis-Ojeda, R., Rappaport, S., Winn, J.~N., {et~al.} 2013, The Astrophysical
  Journal, 774, 54

\bibitem[{{Sanchis-Ojeda} {et~al.}(2015){Sanchis-Ojeda}, {Winn}, {Dai},
  {Howard}, {Isaacson}, {Marcy}, {Petigura}, {Sinukoff}, {Weiss}, {Albrecht},
  {Hirano}, \& {Rogers}}]{Sanchis-Ojeda2015}
{Sanchis-Ojeda}, R., {Winn}, J.~N., {Dai}, F., {et~al.} 2015, \apjl, 812, L11

\bibitem[{{Seager} {et~al.}(2007){Seager}, {Kuchner}, {Hier-Majumder}, \&
  {Militzer}}]{Seager2007}
{Seager}, S., {Kuchner}, M., {Hier-Majumder}, C.~A., \& {Militzer}, B. 2007,
  \apj, 669, 1279

\bibitem[{{Sinukoff} {et~al.}(2017){Sinukoff}, {Howard}, {Petigura}, {Fulton},
  {Isaacson}, {Weiss}, {Brewer}, {Hansen}, {Hirsch}, {Christiansen}, {Crepp},
  {Crossfield}, {Schlieder}, {Ciardi}, {Beichman}, {Knutson}, {Benneke},
  {Dressing}, {Livingston}, {Deck}, {L{\'e}pine}, \&
  {Rogers}}]{Sinukoff2017_W47}
{Sinukoff}, E., {Howard}, A.~W., {Petigura}, E.~A., {et~al.} 2017, \aj, 153, 70

\bibitem[{{Steffen} {et~al.}(2012){Steffen}, {Ragozzine}, {Fabrycky}, {Carter},
  {Ford}, {Holman}, {Rowe}, {Welsh}, {Borucki}, {Boss}, {Ciardi}, \&
  {Quinn}}]{Steffen2012}
{Steffen}, J.~H., {Ragozzine}, D., {Fabrycky}, D.~C., {et~al.} 2012,
  Proceedings of the National Academy of Science, 109, 7982

\bibitem[{{Tsiganis} {et~al.}(2005){Tsiganis}, {Gomes}, {Morbidelli}, \&
  {Levison}}]{Tsiganis2005}
{Tsiganis}, K., {Gomes}, R., {Morbidelli}, A., \& {Levison}, H.~F. 2005, \nat,
  435, 459

\bibitem[{{Venturini} {et~al.}(2015){Venturini}, {Alibert}, {Benz}, \&
  {Ikoma}}]{Venturini2015}
{Venturini}, J., {Alibert}, Y., {Benz}, W., \& {Ikoma}, M. 2015, \aap, 576,
  A114

\bibitem[{{Walsh} {et~al.}(2011){Walsh}, {Morbidelli}, {Raymond}, {O'Brien}, \&
  {Mandell}}]{Walsh2011}
{Walsh}, K.~J., {Morbidelli}, A., {Raymond}, S.~N., {O'Brien}, D.~P., \&
  {Mandell}, A.~M. 2011, \nat, 475, 206

\bibitem[{Weiss \& Marcy(2014)}]{Weiss2014}
Weiss, L.~M., \& Marcy, G.~W. 2014, \apj, 783, L6

\end{thebibliography}

\end{document}